\documentclass[journal]{IEEEtran}
\IEEEoverridecommandlockouts
\usepackage{cite}
\usepackage{amsmath,amssymb,amsfonts}
\usepackage{amsthm}
\usepackage{bbm}
\usepackage{algorithmic}
\usepackage{mathrsfs}
\usepackage[linesnumbered,ruled,vlined]{algorithm2e}
\usepackage{graphicx}
\usepackage{textcomp}
\usepackage{xcolor}
\usepackage{dsfont}
\usepackage{array}
\usepackage{stfloats}
\usepackage{url}
\usepackage{verbatim}
\usepackage{doi}
\usepackage{cite}
\usepackage{amsthm}
\usepackage{graphicx}
\usepackage{subfigure}

\def\BibTeX{{\rm B\kern-.05em{\sc i\kern-.025em b}\kern-.08em
    T\kern-.1667em\lower.7ex\hbox{E}\kern-.125emX}}
\begin{document}

\title{Cooperative Sensing and Uploading for Quality-Cost Tradeoff of Digital Twins in VEC\\
\author{Kai Liu,~\IEEEmembership{Senior Member,~IEEE,} Xincao Xu, Penglin Dai,~\IEEEmembership{Member,~IEEE,} and Biwen Chen
\thanks{Manuscript received. (Corresponding author: Biwen Chen)}
\thanks{Kai Liu, Xincao Xu, and Biwen Chen are with the College of Computer Science, Chongqing University, Chongqing 400040, China. (e-mail: liukai0807, near, macrochen@cqu.edu.cn).}
\thanks{Penglin Dai is with the School of Computing and Artificial Intelligence, Southwest Jiaotong University, Chengdu 611756, China. (e-mail: penglindai@swjtu.edu.cn).}
}
}

\markboth{Journal of \LaTeX\ Class Files,~Vol.~14, No.~8, August~2021}%
{Shell \MakeLowercase{\textit{et al.}}: A Sample Article Using IEEEtran.cls for IEEE Journals}


\maketitle

\begin{abstract}
Recent advances in sensing technologies, wireless communications, and computing paradigms drive the evolution of vehicles in becoming an intelligent and electronic consumer products. This paper investigates enabling digital twins in vehicular edge computing (DT-VEC) via cooperative sensing and uploading, and makes the first attempt to achieve the quality-cost tradeoff in DT-VEC. First, a DT-VEC architecture is presented, where the heterogeneous information can be sensed by vehicles and uploaded to the edge node via vehicle-to-infrastructure (V2I) communications. The digital twins are modeled based on the sensed information, which are utilized to from the logical view to reflect the real-time status of the physical vehicular environment. Second, we derive the cooperative sensing model and the V2I uploading model by considering the timeliness and consistency of digital twins, and the redundancy, sensing cost, and transmission cost. On this basis, a bi-objective problem is formulated to maximize the system quality and minimize the system cost. Third, we propose a solution based on multi-agent multi-objective (MAMO) deep reinforcement learning, where a dueling critic network is proposed to evaluate the agent action based on the value of state and the advantage of action. Finally, we give a comprehensive performance evaluation, demonstrating the superiority of MAMO.
\end{abstract}


\section{Introduction} \label{Intro}
\IEEEPARstart{R}{ecent} advances in sensing technologies, wireless communications, and computing paradigms drive the development of modern new-energy and intelligent vehicles, which are becoming typical intelligent and electronic consumer products.
Various onboard sensors are equipped in modern vehicles to enhance their environment-sensing ability \cite{zhu2017overview}.
On the other hand, the development of vehicle-to-everything (V2X) \cite{chen2020a} enables the cooperation among vehicles, roadside infrastructure and the cloud.
Meanwhile, vehicular edge computing (VEC) \cite{dai2021edge} is a promising paradigm for enabling computation-intensive and latency-critical intelligent transportation system (ITS) \cite{liu2019hierarchical} applications.
These advances become strong driving forces of developing digital twins in vehicular edge computing (DT-VEC).
Specifically, the logical mapping of the physical entities in vehicular networks, such as vehicles, pedestrians and roadside infrastructures, can be constructed at the edge node via the cooperative sensing and uploading.

Great efforts have been devoted to vehicular data dissemination via V2X communications, such as end-edge-cloud cooperative data dissemination architectures \cite{liu2020fog} and intent-based network control frameworks \cite{singh2020intent}.
To improve caching efficiency, some studies proposed vehicular content caching frameworks, such as blockchain-empowered content caching \cite{dai2020deep}, cooperative coding and caching scheduling \cite{xiao2021cooperative}, and edge-cooperative content caching \cite{su2018edge}.
A lot of work have been studied on task offloading in vehicular networks, such as deep-learning-based energy-efficient task offloading \cite{shang2021deep}, real-time multi-period task offloading  \cite{liu2021rtds}, alternating direction method of multipliers (ADMMs) and particle swarm optimization (PSO) combined task offloading \cite{liu2022a}.
These studies mainly focused on scheduling algorithms for data dissemination, information caching, and task offloading in vehicular networks. However, none of them have considered the cooperative sensing and uploading in VEC.

Detection, prediction, planning and control technologies have been also widely studied to enable digital twins in VEC.
A number of detecting technologies have been proposed, such as raindrop quantity detection \cite{wang2021deep} and driver fatigue detection \cite{chang2018design}.
Different methods have been proposed for predicting vehicle status, such as hybrid velocity-profile prediction \cite{zhang2019cyber}, vehicle tracking \cite{iepure2021a} and acceleration prediction \cite{zhang2020data}.
Meanwhile, different scheduling schemes have been proposed in vehicular networks, such as physical-ratio-K interference model-based broadcast scheduling \cite{li2020cyber} and established map model-based path planning \cite{lian2020cyber}.
Some other studies have been focused on controlling algorithms for intelligent vehicles, such as vehicle acceleration controlling \cite{lv2018driving}, intersection controlling \cite{chang2021an} and electric vehicle (EV) charging scheduling \cite{wi2013electric}.
The above studies on status detection, trajectory prediction, path scheduling and vehicle controlling facilitated the implementation of various ITS applications. Nevertheless, they assume the availability of quality information can be constructed in VEC, without considering the sensing and uploading cost.

A few studies have considered the information quality in VCPS, including timeliness \cite{liu2014temporal, dai2019temporal} and accuracy \cite{rager2017scalability, yoon2021performance}.
Nevertheless, they are not sufficient for evaluating the the quality of heterogeneous information fusion in DT-VEC. Several studies considered enabling digital twins in VEC, including edge management framework \cite{zhang2022adaptive}, QoS optimization \cite{xu2022service}, edge caching system \cite{zhang2022digital}. However, none of them have investigated the cooperative sensing and uploading for enabling quality information fusion in DT-VEC.

With above motivation, this paper makes the first attempt to strike the best balance on enhancing the quality of information fusion and minimizing the cost of cooperative sensing and uploading in DT-VEC.
The primary challenges are discussed as follows.
First, due to the highly dynamic information in vehicular networks, it is crucial to evaluate the interrelationship among sensing frequency, queuing delay and transmission delay to enhance information freshness. 
Second, it is possible that redundant or inconsistent information are sensed by different vehicles across different temporal or spatial domains.
Thus, vehicles with different sensing capacities are expected to be cooperated in a distributed way to enhance the utilization of sensing and communication resources. 
Third, the physical information is heterogeneous regarding to distribution, updating frequency and modality, which poses significant difficulties in modeling the quality of information fusion. 
Fourth, higher quality of constructed digital twins comes with higher cost of sensing and communication resources. 
To sum up, realizing the high-quality and low-cost DT-VEC via cooperative sensing and uploading is crucial yet challenging.

The primary contributions of this paper are summarized as follows. 
\begin{itemize}
\item We investigate the problem of distributed information sensing and cooperative V2I uploading in enabling digital twins in VEC. In particular, we design a cooperative sensing model and a V2I uploading model. On this basis, we design the system quality metric to measure the timeliness and consistency of information fusion in DT-VEC and the system cost metric by integrating the redundancy, sensing cost, and transmission cost. Finally, we formulate a bi-objective problem of maximizing the system quality and minimizing the system cost.
\item We propose a solution based on multi-agent multi-objective (MAMO) deep reinforcement learning, which is implemented distributedly in vehicles and the edge node. The action space of each vehicle consists of the sensing decision, sensing frequency, uploading priority and transmission power allocations. The action space of each edge node is the potential V2I bandwidth allocation strategies. The reward vector consists of the individual reward of vehicles via difference reward critic assignment, and normalized reward of the edge node via min-max normalization. A dueling critic network is designed to evaluate agent action based on the value of state and the advantage of action.
\item We build a simulation model based on real-world vehicular trajectories and implement the proposed solution together with three comparable algorithms, including random allocation (RA), distributed distributional deep deterministic policy gradient (D4PG) \cite{barth2018distributed}, and multi-agent D4PG (MAD4PG) \cite{xu2022joint}. We design two new metrics, i.e., quality per unit cost (QPUC) and profit per unit quality (PPUQ) to quantitatively measure the tradeoff achieved by the algorithm. A comprehensive simulation result conclusively demonstrates that the proposed MAMO solution outperforms RA, D4PG and MAD4PG by around 498.8\%, 113.6\% and 451.9\%, respectively, in terms of maximizing the QPUC, and improves the PPUQ by around 34.2\%, 18.6\% and 43.9\% over RA, D4PG and MAD4PG, respectively.
\end{itemize}

The rest of this paper is organized as follows.
Section \ref{architecture} presents the system architecture.
Section \ref{model} develops the system model.
Section \ref{problem} formulates the problem.
Section \ref{solution} proposes the solution.
Section \ref{evaluation} evaluates the performance. 
Finally, Section \ref{conclusion} concludes this paper and discusses future research directions.

\section{DT-VEC Architecture} \label{architecture}

\begin{figure}
\centering
  \includegraphics[width=0.99\columnwidth]{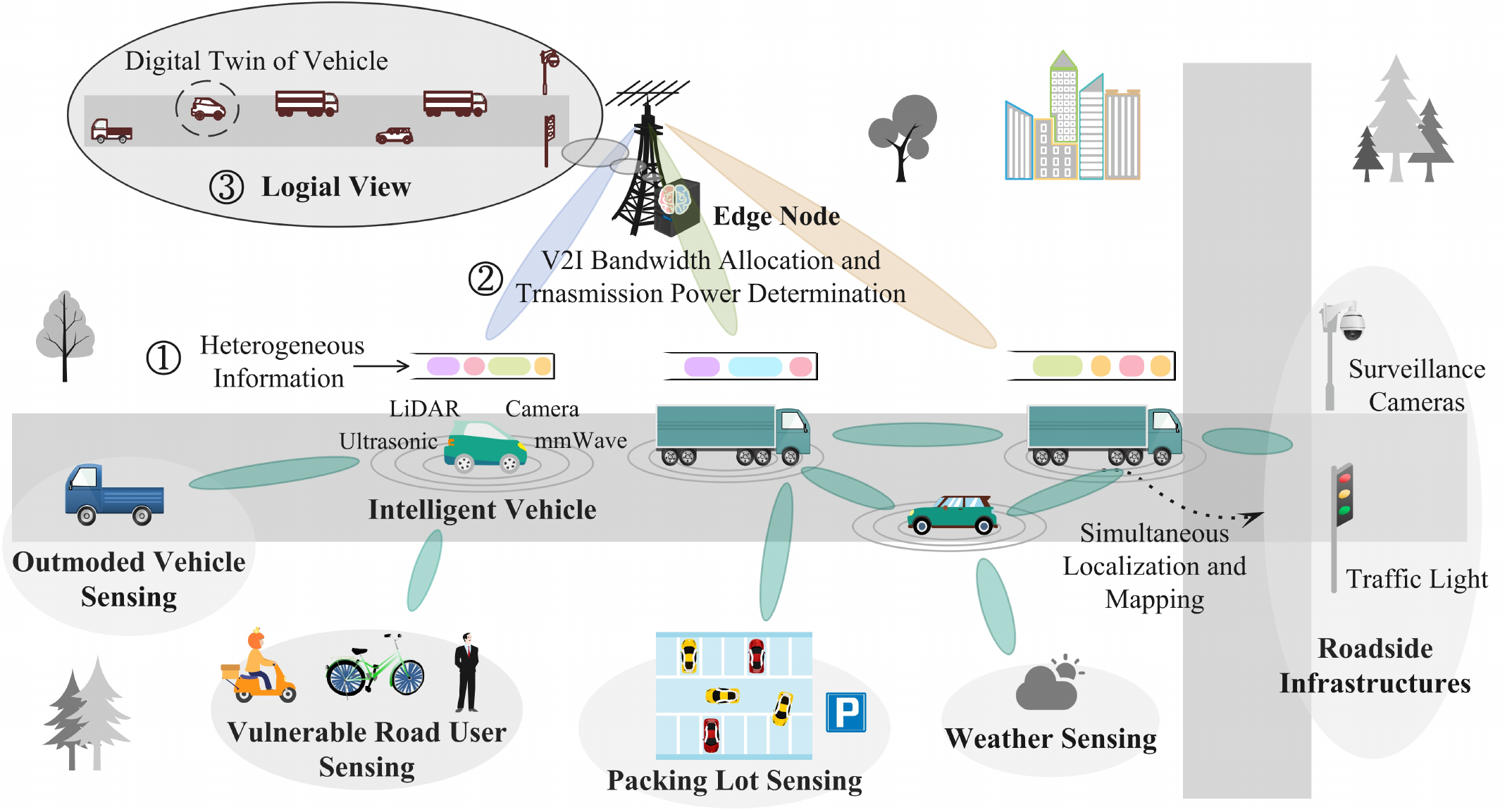}
  \caption{DT-VEC architecture}
  \label{fig_0_system_example}
\end{figure} 

In this section, we present the DT-VEC architecture as shown in Fig. \ref{fig_0_system_example}.
Vehicles are equipped with various onboard sensors, such as ultrasonic sensors, LiDAR, optical cameras and mmWave radars.
The heterogeneous information, including the status of vehicles, road users, packing lots and roadside infrastructures can collaboratively sensed by vehicles via onboard sensors.
Such information can be used to model the digital twins in the edge node, which can be further utilized to from the logical views and enable various ITS applications, such as simultaneous localization and mapping (SLAM) and autonomous intersection control system, with requirements of the fusion of different modalities of information of physical entities in vehicular networks.
A high-quality of logical views can better reflect the real-time physical vehicular environments, and thus enhance the ITS performances. 
Nevertheless, constructing high-quality of logical views may come with of higher sensing frequency, more amount of information uploading and higher energy consumption.

The system workflow is described as follows. 
First, the real-time status of different physical entities are sensed by vehicles and queued up for uploading. 
Second, the edge node allocates V2I bandwidth to vehicles, and on this basis, each vehicle determine the transmission power. 
Third, the digital twins of physical entities are modeled based on the fusion of heterogeneous information received from vehicles.
Note that in such a system, the heterogeneous information are sensed by vehicles with different sensing frequencies, resulting in different freshness when uploading. 
Although increasing the sensing frequency may improve the freshness, it may prolong the queuing latency and cost more energy.
In addition, the information of specific physical entity may be sensed by multiple vehicles, which may waste the communication resources if uploading by all vehicles. 
Then, due to the constraints of transmission power and V2I bandwidth, it is important to allocate the communication resources efficiently and economically to improve resource utilization.
As mentioned above, it is critical and challenging to quantitatively measure the quality and cost of digital twins  constructed at edge nodes, and design an efficient and economical scheduling mechanism for cooperative sensing and uploading to maximize the system quality and minimize the system cost in DT-VEC.

\section{System Model} \label{model}
\subsection{Notations}
The set of discrete time slots is denoted by $\mathbf{T}=\left\{1,2,\cdots,t,\cdots, T \right\}$, where $T$ is the number of time slots.
Let $\mathbf{D}$ denote the set of heterogeneous information, and each information $d \in \mathbf{D}$ is characterized by a three-tuple $d=\left(\operatorname{type}_d, u_d, \left|d\right| \right)$, where $\operatorname{type}_d$, $u_d$ and $\left|d\right|$ are the type, updating interval and data size, respectively. 
We denote $\mathbf{S}$ as the set of vehicles, and each vehicle $s \in \mathbf{S}$ is characterized by a three-tuple $s=\left (l_s^t, \mathbf{D}_s, \pi_s \right )$, where $l_s^t$, $\mathbf{D}_s$ and $\pi_s$ are the location, sensed information set and transmission power, respectively.
For each information $d \in \mathbf{D}_s$, the sensing cost (i.e., energy consumption) in vehicle $s$ is denoted by $\phi_{d, s}$.
Let $\mathbf{E}$ denote the set of edge nodes.
Each edge node $e \in \mathbf{E}$ is characterized by a three-tuple $e=\left (l_e, r_e, b_e \right)$, where $l_{e}$, $r_{e}$ and $b_{e}$ are the location, communication range and bandwidth, respectively.
The distance between vehicle $s$ and edge node $e$ is denoted by $\operatorname{dis}_{s,e}^t \triangleq \operatorname{distance} \left (l_s^t,l_e \right ), \forall s \in \mathbf{S}, \forall e \in \mathbf{E}, \forall t \in \mathbf{T}$, where $\operatorname{distance}\left(\cdot,\cdot\right)$ is the Euclidean distance.
The set of vehicles within the radio coverage of edge node $e$ at time $t$ is denoted by $\mathbf{S}_e^t=\left \{s \vert \operatorname{dis}_{s,e}^t \leq r_e, \forall s \in \mathbf{S} \right \}, \mathbf{S}_e^t \subseteq \mathbf{S}$.

The sensing decision indicator, indicating whether information $d$ is sensed by vehicle $s$ at time $t$, is denoted by
\begin{equation}
	c_{d, s}^t \in \{0, 1\}, \forall d \in \mathbf{D}_{s}, \forall s \in \mathbf{S}, \forall t \in \mathbf{T}
	\label{constrain 1} 
\end{equation}
Then, the set of information sensed by vehicle $s$ at time $t$ is denoted by $\mathbf{D}_s^t = \{ d | c_{d, s}^{t} = 1, \forall d \in \mathbf{D}_s \}, \mathbf{D}_s^t \subseteq \mathbf{D}_s$.
The information types are distinct for any information $d \in \mathbf{D}_s^t$, i.e., $\operatorname{type}_{d^*} \neq \operatorname{type}_{d}, \forall d^* \in \mathbf{D}_s^t \setminus \left\{ d\right \}, \forall d \in \mathbf{D}_s^t$.
The sensing frequency of information $d$ in vehicle $s$ at time $t$ is denoted by $\lambda_{d,s}^t$, which should meet the requirement of the sensing ability of vehicle $s$.
\begin{equation}
	\lambda_{d,s}^{t} \in [\lambda_{d,s}^{\min} , \lambda_{d,s}^{\max} ], \ \forall d \in \mathbf{D}_s^t, \forall s \in \mathbf{S}, \forall t \in \mathbf{T}
	\label{constrain 2} 
\end{equation}
where $\lambda_{d,s}^{\min}$ and $\lambda_{d,s}^{\max}$ are the minimum and maximum of sensing frequency for information ${d}$ in vehicle $s$, respectively.
The uploading priority of information $d$ in vehicle $s$ at time $t$ is denoted by $p_{d,s}^t$, and we have
\begin{equation}
	{p}_{d^*, s}^t \neq {p}_{d, s}^t, \forall d^* \in \mathbf{D}_s^t \setminus \left\{ d\right \}, \forall d \in \mathbf{D}_s^t, \forall s \in \mathbf{S}, \forall t \in \mathbf{T}
	\label{constrain 3} 
\end{equation}
where ${p}_{d^*, s}^t$ is the uploading priority of information $d^* \in \mathbf{D}_s^t$.
The transmission power of vehicle $s$ at time $t$ is denoted by $\pi_{s}^t$, and it can not exceed the power capacity of vehicle $s$. 
\begin{equation}
	\pi_s^t \in \left[ 0 , \pi_s \right ], \forall s \in \mathbf{S}, \forall t \in \mathbf{T}
	\label{constrain 4} 
\end{equation}
The V2I bandwidth allocated by edge node $e$ for vehicle $s$ at time $t$ is denoted by $b_{s, e}^t$, and we have 
\begin{equation}
	b_{s, e}^t \in \left [0, b_e \right], \forall s \in \mathbf{S}_e^{t}, \forall e \in \mathbf{E}, \forall t \in \mathbf{T}
	\label{constrain 5} 
\end{equation}
The total V2I bandwidth allocated by edge node $e$ cannot exceed its capacity $b_e$, i.e., ${\sum_{\forall s \in \mathbf{S}_e^{t}} b_{s, e}^t} \leq b_e, \forall t \in \mathbf{T}$.

\begin{table}[ht]\footnotetext
\centering
\caption{Summary of primary notations}
\begin{tabular}[t]{|p{0.05\textwidth}<{\centering}||p{0.40\textwidth}<{\raggedright}|}
\hline
Notations&Descriptions\\
\hline
\hline
$\mathbf{T}$, $\mathbf{D}$&Set of discrete time slots, Set of heterogeneous information\\
\hline
$\mathbf{S}$, $\mathbf{E}$&Set of vehicles, Set of edge nodes\\
\hline
$\operatorname{type}_d$&Type of information $d$\\
\hline
$u_d$&Updating interval of information $d$\\
\hline
$l_{s}^{t}$&Location of vehicle $s$ at time $t$ \\
\hline
$\mathbf{D}_s$&Set of information sensed by vehicle $s$\\
\hline
$\pi_{s}$&Transmission power of vehicle $s$\\
\hline
$l_{e}$&Location of edge node $e$\\
\hline
$r_{e}$&Communication range of edge node $e$\\
\hline
$b_{e}$&Bandwidth capacity of edge node $e$\\
\hline
$\operatorname{dis}_{s, e}^{t}$&Distance between vehicle $s$ and edge node $e$ at time $t$\\
\hline
$S_{e}^{t}$&Set of vehicles within the coverage of edge node $e$\\
\hline
$\mathbf{V}^{\prime}$&Set of physical entities\\
\hline
$\mathbf{V}$&Set of digital twins\\
\hline
$\mathbf{D}_{v^{\prime}}$&Set of information associated with physical entity $v^{\prime}$\\
\hline
$\mathbf{V}_e^t$&Set of digital twins modeled in edge node $e$ at time $t$\\
\hline
$\mathbf{D}_{v, e}^t$&Set of information received by edge node $e$ and required by $v$\\
\hline
$c_{d, s}^t$&Binary indicates whether information $d$ is sensed by vehicle $s$\\
\hline
$\mathbf{D}_s^t$&Set of information sensed by vehicle $s$ at time $t$\\
\hline
$\lambda_{d, s}^{t}$&Sensing frequency of information $d$ in vehicle $s$ at time $t$\\
\hline
$p_{d, s}^{t}$&Uploading priority of information $d$ in vehicle $s$ at time $t$\\
\hline
$\pi_{s}^t$&Transmission power of vehicle $s$ at time $t$\\
\hline
$b_{s, e}^{t}$&Bandwidth allocated by edge node $e$ for vehicle $s$ at time $t$\\
\hline
$\operatorname{a}_{d,s}^t$&Arrival moment of the information with $\operatorname{type}_d$ in vehicle $s$\\
\hline
$\operatorname{u}_{d,s}^t$&Updating moment of the information with $\operatorname{type}_d$ in vehicle $s$\\
\hline
$\mathbf{D}_{d, s}^t$&Set of information with higher uploading priority than $d$\\
\hline
$\operatorname{q}_{d, s}^t$&Queuing time of information $d$ in vehicle $s$\\
\hline
$\operatorname{g}_{d, s, e}^t$&Transmission time of information $d$\\
\hline
$\mathbf{D}_{s, e}^t$&Set of information transmitted by vehicle $s$ and received at $e$\\ 
\hline
\end{tabular}
\label{table_notations}
\end{table}

Denote the physical entity, such as vehicle, pedestrian, and roadside infrastructure, by $v^{\prime}$, and denote the set of physical entities as $\mathbf{V}^{\prime}$ in VEC.
$\mathbf{D}_{v^{\prime}}$ is the set of information associated with entity $v^{\prime}$, and can be represented by $\mathbf{D}_{v^{\prime}}=\left\{d \mid y_{d,v^{\prime}} = 1, \forall d \in \mathbf{D} \right\}, \forall v^{\prime} \in \mathbf{V}^{\prime}$, where $y_{d, v^{\prime}}$ is a binary indicating whether information $d$ is associated by entity $v^{\prime}$.
The size of $\mathbf{D}_{v^{\prime}}$ is denoted by $|\mathbf{D}_{v^{\prime}}|$.
Each entity may require multiple pieces of information, i.e., $|\mathbf{D}_{v^{\prime}}| = \sum_{\forall d \in \mathbf{D}}y_{d, v^{\prime}} \geq 1, \forall v^{\prime} \in \mathbf{V}^{\prime}$.
For each entity $v^{\prime} \in \mathbf{V}^{\prime}$, there may be a digital twin $v$ modeled in the edge node.
We denote the set of digital twins by $V$,  and the set of digital twins modeled in edge node $e$ at time $t$ is denoted by $\mathbf{V}_e^{t}$.
Therefore, the set of information received by edge node $e$ and required by digital twin $v$ can be represented by $\mathbf{D}_{v, e}^t=\bigcup_{\forall s \in \mathbf{S}}\left(\mathbf{D}_{v^{\prime}} \cap \mathbf{D}_{s, e}^t\right), \forall v \in \mathbf{V}_e^{t}, \forall e \in \mathbf{E}$, and $| \mathbf{D}_{v,e}^t |$ is the number of information received by edge node $e$ and required by digital twin $v$,  which is computed by $| \mathbf{D}_{v,e}^t | =  \sum_{\forall s \in \mathbf{S}} \sum_{\forall d \in \mathbf{D}_s} c_{d, s}^t  y_{d, v^{\prime}}$.
The primary notations are summarized in Table \ref{table_notations}.

\theoremstyle{definition}
\newtheorem{Definition}{Definition}

\subsection{Cooperative Sensing Model}
The cooperative sensing is modeled based on multi-class M/G/1 priority queue \cite{moltafet2020age}.
Assume the uploading time $\operatorname{\hat{g}}_{d, s, e}^t$ of information with $\operatorname{type}_d$ follows a class of General distribution with mean $\alpha_{d, s}^t$ and variance $\beta_{d, s}^t$.
Then, the uploading workload $\rho_{s}^{t}$ in vehicle $s$ is represented by $ \rho_{s}^{t}=\sum_{\forall d \subseteq \mathbf{D}_s^t} \lambda_{d,s}^{t} \alpha_{d, s}^t$.
According to the principle of multi-class M/G/1 priority queue, it requires $\rho_{s}^{t} < 1$ to have the steady-state of the queue.
The arrival time of information $d$ before time $t$ is denoted by $\operatorname{a}_{d,s}^t$, which is computed by
\begin{equation}
    \operatorname{a}_{d,s}^t = { {1}/{\lambda_{d, s}^{t}} \left \lfloor t \lambda_{d,s}^t \right \rfloor} 
\end{equation}
The updating time of information $d$ before time $t$ denoted by $\operatorname{u}_{d,s}^t$ is computed by
\begin{equation}
    \operatorname{u}_{d,s}^t = \left \lfloor  {\operatorname{a}_{d,s}^t}/{u_d} \right \rfloor  u_d
\end{equation}
where $u_d$ is the updating interval of information $d$.

The set of elements with higher uploading priority than $d$ in vehicle $s$ at time $t$ is denoted by $\mathbf{D}_{d, s}^t = \{ d^* \mid p_{d^*,s}^{t} > p_{d,s}^{t} , \forall d^* \in \mathbf{D}_s^t \}$, where $p_{d^*,s}^{t}$ is the uploading priority of information $d^* \in \mathbf{D}_s^t$.
Thus, the uploading workload ahead of information $d$ (i.e., the amount of elements to be uploaded before $d$ by vehicle $s$ at time $t$) is computed by 
\begin{equation}
	\rho_{d, s}^{t}=\sum_{\forall d^* \in \mathbf{D}_{d, s}^t} \lambda_{d^*,s}^t \alpha_{d^*, s}^t
\end{equation}
where $\lambda_{d^*,s}^t$ and $\alpha_{d^*, s}^t$ are the sensing frequency and the mean transmission time of information $d^*$ in vehicle $s$ at time $t$, respectively.
According to the Pollaczek-Khintchine formula \cite{takine2001queue}, the queuing time of information $d$ in vehicle $s$ is calculated by 
\begin{equation}
    \text{\footnotesize$\operatorname{q}_{d, s}^t= \frac{1} {1 - \rho_{d,s}^{t}} 
        \left[ \alpha_{d, s}^t + \frac{ \lambda_{d,s}^{t} \beta_{d, s}^t + \sum\limits_{\forall d^* \in \mathbf{D}_{d, s}^t} \lambda_{d^*,s}^t \beta_{d^*, s}^t }{2\left(1-\rho_{d,s}^{t} - \lambda_{d,s}^{t} \alpha_{d, s}^t\right)}\right] 
        - \alpha_{d, s}^t$}
\end{equation}

\subsection{V2I Uploading Model}
The V2I uploading is modeled based on channel fading distribution and SNR threshold.
The SNR of V2I communications between vehicle $s$ and edge node $e$ at time $t$ is computed by \cite{sadek2009distributed}
\begin{equation}
    \label{equ_SNR}
    \operatorname{SNR}_{s,e}^{t}=\frac{1}{N_{0}} \left|h_{s,e}\right|^{2} \tau {\operatorname{dis}_{s,e}^{t}}^{-\varphi} {\pi}_s^t
\end{equation}
where $N_{0}$ is the additive white Gaussian noise; $h_{s, e}$ is the channel fading gain; $\tau$ is a constant that depends on the antennas design, and $\varphi$ is the path loss exponent.
Assume that $\left|h_{s,e}\right|^{2}$ follows a class of distributions with the mean $\mu_{s,e}$ and variance $\sigma_{s,e}$, which is represented by
\begin{equation}
    \tilde{p}=\left\{\mathbb{P}: \mathbb{E}_{\mathbb{P}}\left[\left|h_{s, e}\right|^{2}\right]=\mu_{s,e}, \mathbb{E}_{\mathbb{P}}\left[\left|h_{s, e}\right|^{2}-\mu_{s,e}\right]^{2}=\sigma_{s,e}\right\}
\end{equation}
The transmission reliability is measured by the possibility that a successful transmission probability is beyond a reliability threshold.
\begin{equation}
    \inf_{\mathbb{P} \in \tilde{p}} \operatorname{Pr}_{[\mathbb{P}]}\left(\operatorname{SNR}_{s,e}^{t} \geq \operatorname{SNR}_{s,e}^{\operatorname{tgt}}\right) \geq \delta
\end{equation}
\noindent where $\operatorname{SNR}_{s,e}^{\operatorname{tgt}}$ and $\delta$ are the target SNR threshold and reliability threshold, respectively.
The set of information uploaded by vehicle $s$ and received by edge node $e$ is denoted by $\mathbf{D}_{s, e}^{t} = \bigcup_{\forall s \in \mathbf{S}_{e}^{t}} \mathbf{D}_{s}^{t}$.

According to the Shannon theory, the transmission rate of V2I communications between vehicle $s$ and edge node $e$ at time $t$ denoted by $\operatorname{z}_{s,e}^t$, is computed by 
\begin{equation}
    \operatorname{z}_{s,e}^t=b_{s}^{t} \log _{2}\left(1+\mathrm{SNR}_{s, e}^{t}\right)
\end{equation}
Thus, the duration for transmitting information $d$ from vehicle $s$ to edge node $e$, denoted by $\operatorname{g}_{d, s, e}^t$, is computed by 
\begin{equation}
    \operatorname{g}_{d, s, e}^t=\inf _{j \in \mathbb{R}^+} \left \{ \int_{\operatorname{k}_{d, s}^t}^{\operatorname{k}_{d, s}^t + j} {\operatorname{z}_{s,e}^t} \operatorname{d}t \geq \left|d\right| \right \} 
\end{equation}
\noindent where $\operatorname{k}_{d, s}^t = t +\operatorname{q}_{d, s}^t$ is the moment when vehicle $s$ starts to transmit information $d$.

\section{Problem Formulation} \label{problem}

\subsection{Quality of Digital Twin}
First, since the digital twins are modeled based on the continuously uploaded and time-varying information, we define the timeliness of information $d$ as follows.
\begin{Definition}[\textit{Timeliness of information $d$}]
The timeliness $\theta_{d,s} \in \mathbb{Q}^{+}$ of information $d$ in vehicle $s$ is defined as the duration between the updating and the receiving  of information $d$.
\begin{equation}
    \theta_{d,s} = \operatorname{a}_{d,s}^t + \operatorname{q}_{d,s}^t + \operatorname{g}_{d, s, e}^t-\operatorname{u}_{d,s}^{t}, \forall d \in \mathbf{D}_s^t,\forall s \in \mathbf{S}
\end{equation}
\end{Definition}
\begin{Definition}[\textit{Timeliness of digital twin $v$}]
	The timeliness $\Theta_{v} \in \mathbb{Q}^{+}$ of digital twin $v$ is defined as the sum of the maximum timeliness of information associated with physical entity $v^{\prime}$.
	\begin{equation}
    	\Theta_{v} = \sum_{\forall s\in \mathbf{S}_{e}^{t}} \max_{\forall d \in \mathbf{D}_{v^{\prime}} \cap \mathbf{D}_s^t}\theta_{d,s}, \forall v \in \mathbf{V}_{e}^{t}, \forall e \in \mathbf{E}
    	\label{definition of timeliness}
	\end{equation}
\end{Definition}

Second, as different types of information have different sensing frequencies and uploading priorities, we define the consistency of digital twin to measure the consistency of information  associated with the same physical entity. 
\begin{Definition}[\textit{Consistency of digital twin $v$}]
	 The consistency $\Psi_{v} \in \mathbb{Q}^{+}$ of digital twin $v$ is defined as the maximum of the difference between information updating time.
\begin{equation}
    \Psi_{v}=\max_{\forall d \in \mathbf{D}_{v, e}^{t}, \forall s \in \mathbf{S}_{e}^{t}} {\operatorname{u}_{d,s}^t} - \min_{\forall d \in \mathbf{D}_{v, e}^{t}, \forall s \in \mathbf{S}_{e}^{t}} {\operatorname{u}_{d,s}^t} , \forall v \in \mathbf{V}_{e}^{t}, \forall e \in \mathbf{E}
    \label{definition of consistency}
\end{equation}
\end{Definition}

Finally, we give the formal definition of the quality of digital twin, synthesizing the timeliness and consistency of digital twin.
\begin{Definition}[\textit{Quality of digital twin, QDT}]
	The quality of digital twin $\operatorname{QDT}_{v} \in (0, 1)$ is defined as a weighted average of normalized timeliness and normalized consistency of digital twin $v$.
	\begin{equation}
	    \operatorname{QDT}_{v} = w_1 (1 -\hat{\Theta_{v}}) + w_2 (1 - \hat{\Psi_{v}}), \forall v \in \mathbf{V}_{e}^t, \forall e \in \mathbf{E}
	    \label{definition of AoV}
	\end{equation}
\end{Definition}
\noindent where $\hat{\Theta_{v}} \in (0, 1)$ and $\hat{\Psi_{v}} \in (0, 1)$ denote the normalized timeliness and normalized consistency, respectively, which can be obtained by rescaling the range of the timeliness and consistency in $(0, 1)$ via the min-max normalization.
The weighting factors for $\hat{\Theta_{v}}$ and $\hat{\Psi_{v}}$ are denoted by $w_1$ and $w_2$, respectively, which can be tuned accordingly based on the different requirements of ITS applications and we have $w_1+w_2=1$.

\subsection{Cost of Digital Twin}

First, because the status of the same physical entity might be sensed by multiple vehicles simultaneously, we define the redundancy of information $d$ as follows.
\begin{Definition}[\textit{Redundancy of information $d$}]
	 The redundancy $\xi_d \in \mathbb{N}$ of information $d$ is defined as the number of additional information with $\operatorname{type}_d$ sensed by vehicles.
\begin{equation}
    \xi_d= \left | \mathbf{D}_{d, v, e} \right| - 1, \forall d \in \mathbf{D}_v, \forall v \in \mathbf{V}_{e}^{t}, \forall e \in \mathbf{E}
\end{equation}
\noindent where $\mathbf{D}_{d, v, e}$ is the set of the information with $\operatorname{type}_d$ received by edge node $e$ and  required by digital twin $v$, which is represented by $\mathbf{D}_{d, v, e}=\left\{ d^* \vert \operatorname{type}_{d^*} = \operatorname{type}_{d}, \forall d^* \in \mathbf{D}_{v, e}^t \right \}$.
\end{Definition}
\begin{Definition}[\textit{Redundancy of digital twin $v$}]
	The redundancy $\Xi_v \in \mathbb{N}$ of digital twin $v$ is defined as the total redundancy of information in digital twin $v$.
	\begin{equation}
       \Xi_v =  \sum_{\forall d \in \mathbf{D}_{v^{\prime}}} \xi_d, \forall v \in \mathbf{V}_{e}^{t}, \forall e \in \mathbf{E}
       \label{definition of redundancy}
    \end{equation}
\end{Definition}

Second, information sensing and transmission require energy consumption of vehicles, we define the sensing cost and transmission cost of digital twin $v$ as follows.
\begin{Definition}[\textit{Sensing cost of digital twin $v$}]
	The sensing cost $\Phi_{v} \in \mathbb{Q}^{+}$ of digital twin $v$ is defined as the total sensing cost of information required by digital twin $v$.
	\begin{equation}
        \Phi_{v} = \sum_{\forall s \in \mathbf{S}_{e}^{t}} \sum_{\forall d \in \mathbf{D}_{v^{\prime}} \cap \mathbf{D}_s^t}{\phi_{d,s}}, \forall v \in \mathbf{V}_{e}^t, \forall e \in \mathbf{E}
        \label{definition of sensing cost}
    \end{equation}
    where $\phi_{d,s}$ is the sensing cost of information $d$ in vehicle $s$.
\end{Definition}
\begin{Definition}[\textit{Transmission cost of information $d$}]
The transmission cost ${\omega}_{d,s} \in \mathbb{Q}^{+}$ of information $d$ in vehicle $s$ is defined as the consumed transmission power during the information uploading.
\begin{equation}
    {\omega}_{d,s}= \pi_s^t \operatorname{g}_{d, s, e}^t, \forall d \in \mathbf{D}_s^t
\end{equation}
where $\pi_s^t$ and $\operatorname{g}_{d, s, e}^t$ are the transmission power and transmission time, respectively.
\end{Definition}
\begin{Definition}[\textit{Transmission cost of digital twin $v$}]
	The transmission cost $\Omega_{v} \in \mathbb{Q}^{+}$ of digital twin $v$ is defined as the total transmission cost of information required by digital twin $v$.
	\begin{equation}
        \Omega_{v} = \sum_{\forall s \in \mathbf{S}_{e}^{t}} \sum_{\forall d \in \mathbf{D}_{v^{\prime}} \cap \mathbf{D}_s^t} {\omega}_{d,s}, \forall v \in \mathbf{V}_{e}^t, \forall e \in \mathbf{E}
       	\label{definition of transmission cost}
    \end{equation}
\end{Definition}

Finally, we give the formal definition of the cost of digital twin, synthesizing the redundancy, sensing cost, and transmission cost.
\begin{Definition}[\textit{Cost of digital tiwn, CDT}]
	The cost of digital twin $\operatorname{CDT}_{v} \in (0, 1)$ is defined as a weighted average of normalized redundancy, normalized sensing cost, and normalized transmission cost of digital twin $v$.
	\begin{equation}
	    \operatorname{CDT}_{v} = w_3  \hat{\Xi_{v}} +  w_4 \hat{\Phi_{v}} + w_5 \hat{\Omega_{v}}, \forall v \in \mathbf{V}_{e}^t, \forall e \in \mathbf{E}
	    \label{definition of CoV}
	\end{equation}
\end{Definition}
\noindent where $\hat{\Xi_{v}}\in (0, 1)$, $\hat{\Phi_{v}} \in (0, 1)$ and $\hat{\Omega_{v}} \in (0, 1)$ denote the normalized redundancy, normalized sensing cost and normalized transmission cost of digital twin $v$, respectively.
The weighting factors for $\hat{\Xi_{v}}$, $\hat{\Phi_{v}}$ and $\hat{\Omega_{v}}$ are denoted by $w_3$, $w_4$ and $w_5$, respectively.
Similarly, we have $w_3+w_4+w_5=1$.

\subsection{A Bi-Objective Problem}
We define the system quality and system cost as follows.
\begin{Definition}[\textit{System quality}]
	The system quality $\mathscr{Q} \in (0, 1)$ is defined as the average of QDT for each digital twin modeled in edge nodes during the scheduling period $\mathbf{T}$.
	\begin{equation}
		\mathscr{Q}=\frac{\sum_{\forall t \in \mathbf{T}} \sum_{\forall e \in \mathbf{E}} \sum_{\forall v \in \mathbf{V}_e^t} \operatorname{QDT}_{v}}{\sum_{\forall t \in \mathbf{T}} \sum_{\forall e \in \mathbf{E}} |\mathbf{V}_e^t| }
		\label{definition of system quality}
	\end{equation}
\end{Definition}
\begin{Definition}[\textit{System cost}]
	The system cost $\mathscr{C} \in (0, 1)$ is defined as the average of CDT for each digital twin model in edge nodes during the scheduling period $\mathbf{T}$.
	\begin{equation}
		\mathscr{C}=\frac{\sum_{\forall t \in \mathbf{T}} \sum_{\forall e \in \mathbf{E}} \sum_{\forall v \in \mathbf{V}_e^t}  \operatorname{CDT}_{v}}{\sum_{\forall t \in \mathbf{T}} \sum_{\forall e \in \mathbf{E}} |\mathbf{V}_e^t| }
		\label{definition of system cost}
	\end{equation}
\end{Definition}

Given a solution $x = ( \mathbf{C}, \bf\Lambda, \mathbf{P}, \bf\Pi, \mathbf{B} )$, where $\mathbf{C}$ denotes the determined sensing information, $\bf\Lambda$ denotes the determined sensing frequencies, $\mathbf{P}$ denotes the determined uploading priorities, $\bf\Pi$ denotes the determined transmission power, and $\mathbf{B}$ denotes the determined V2I bandwidth allocation:
\begin{equation}
    \begin{cases}
\mathbf{C} &= \left \{ c_{d, s}^t \vert \forall d \in D_{s}, \forall s \in S, \forall t \in T \right  \} \\
\bf\Lambda &= \left \{ \lambda_{d,s}^{t} \vert \forall d \in D_s^t  , \forall s \in S, \forall t \in T \right \} \\ 
\mathbf{P} &= \left \{ p_{d,s}^{t} \vert \forall d \in D_s^t  , \forall s \in S, \forall t \in T\right \}  \\
\bf\Pi &= \left \{ \pi_s^t \vert \forall s \in S, \forall t \in T \right \} \\
\mathbf{B} &= \left \{ b_s^t \vert \forall s \in S, \forall t \in T\right \}
\end{cases}
\end{equation}
where $c_{d, s}^t$, $\lambda_{d,s}^{t}$ and $p_{d,s}^{t}$ are the sensing decision, sensing frequency and uploading priority of information $d$ in vehicle $s$ at time $t$, respectively, and $\pi_s^t$ and $b_s^t$ are the transmission power and V2I bandwidth of vehicle $s$ at time $t$, respectively.
We formulate the bi-objective problem aiming at maximizing the system quality and minimizing the system cost simultaneously, which is expressed by
\begin{equation}
	\begin{aligned}
		\operatorname{\mathbf{P1}}: & \max_x \mathscr{Q}, \min_x \mathscr{C}\\
		\text { s.t. }
		& (1) \sim (5) \\
        &\operatorname{C1}: \sum_{\forall d \subseteq \mathbf{D}_s^t} \lambda_{d,s}^{t} \mu_d<1,\ \forall s \in \mathbf{S}, \forall t \in \mathbf{T} \\
        &\operatorname{C2}: \inf_{\mathbb{P} \in \tilde{p}} \operatorname{Pr}_{[\mathbb{P}]}\left(\operatorname{SNR}_{s,e}^{t} \geq \operatorname{SNR}_{s,e}^{\operatorname{tgt}}\right) \geq \delta, \forall s \in \mathbf{S}, \forall t \in \mathbf{T}\\
        &\operatorname{C3}: {\sum_{\forall s \in \mathbf{S}_e^{t}}b_s^t} \leq b_e, \forall t \in \mathbf{T}
	\end{aligned}
\end{equation}
where $\operatorname{C1}$ guarantees the queue steady-state and $\operatorname{C2}$ guarantees the transmission reliability.
$\operatorname{C3}$ requires that the sum of V2I bandwidth allocated by the edge node $e$ cannot exceed its capacity $b_e$.
According to the definition of CDT, we define the profit of digital twin as follows.
\begin{Definition}[\textit{Profit of digital twin, PDT}]
	The profit of digital twin $\operatorname{PDT}_{v} \in (0, 1)$ is defined as the complement of CDT of digital twin $v$.
	\begin{equation}
		\mathscr{P}= 1 - \operatorname{CDT}_{v}
	\end{equation}
\end{Definition}
\noindent Then, we define the system profit as follows.
\begin{Definition}[\textit{System profit}]
	The system profit $\mathscr{P} \in (0, 1)$ is defined as the average of the PDT for each digital twin modeled in edge nodes during the scheduling period $\mathbf{T}$.
	\begin{equation}
		\mathscr{P}= \frac{\sum_{\forall t \in \mathbf{T}} \sum_{\forall e \in \mathbf{E}} \sum_{\forall v \in \mathbf{V}_e^t}   \operatorname{PDT}_{v} }{\sum_{\forall t \in \mathbf{T}} \sum_{\forall e \in \mathbf{E}} |\mathbf{V}_e^t| }
		\label{definition of view profit}
	\end{equation}
\end{Definition}
\noindent Thus, the problem $\operatorname{\mathbf{P1}}$ can be rewritten as follows.
\begin{equation}
	\begin{aligned}
		\operatorname{\mathbf{P2}}: & \max_x \left (\mathscr{Q}, \mathscr{P} \right )\\
		\text { s.t. }
		&(1) \sim (5), \operatorname{C1} \sim \operatorname{C3}
	\end{aligned}
\end{equation}

\section{Proposed Solution} \label{solution}

\begin{figure*}
\centering
  \includegraphics[width=1.99\columnwidth]{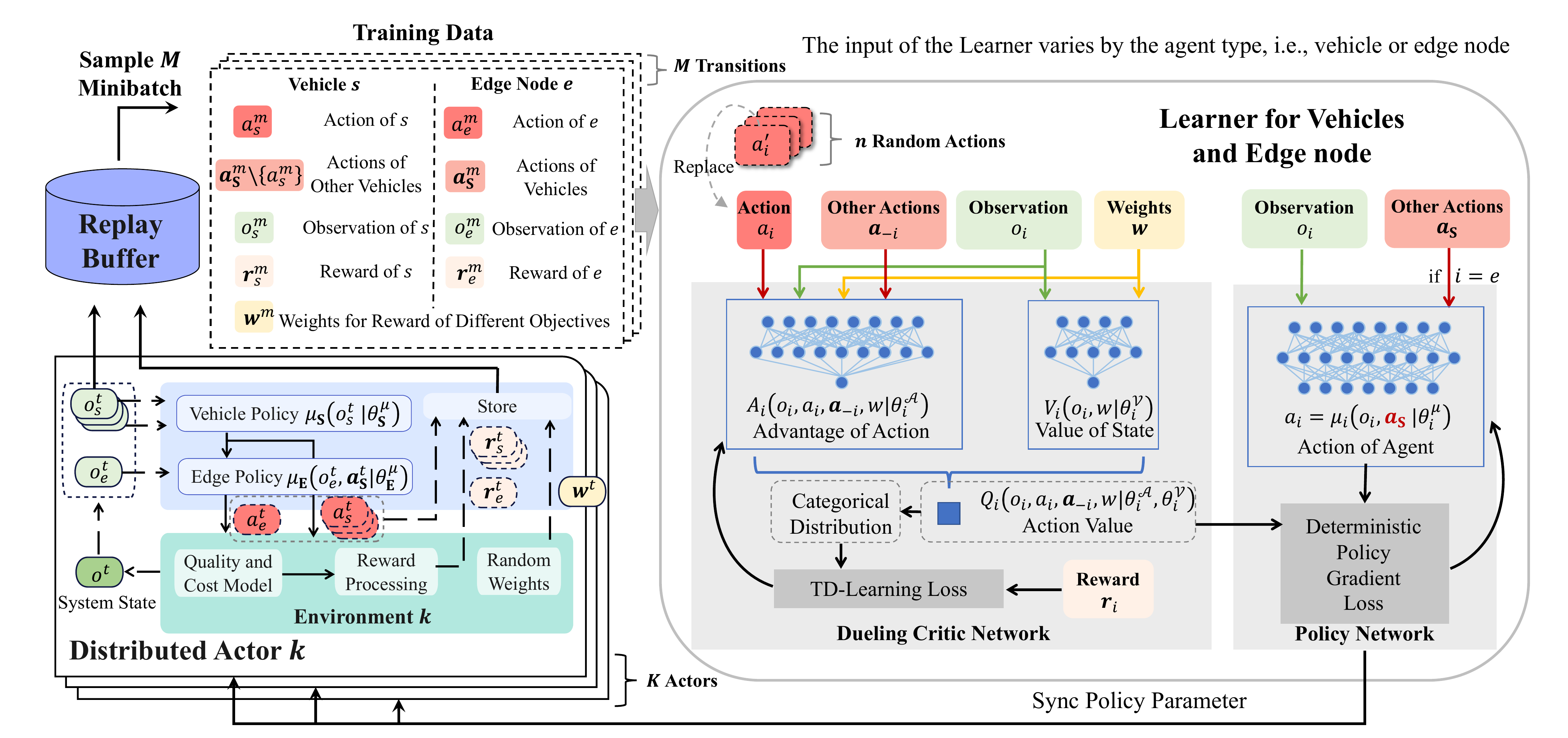}
  \caption{Multi-agent multi-objective deep reinforcement learning model}
  \label{fig_soultion_model}
\end{figure*}

In this section, we propose a solution based on multi-agent multi-objective deep reinforcement learning.
The general model of the solution is shown in Fig. \ref{fig_soultion_model}, which consists of $K$ distributed actors, a learner, and a replay buffer.
Specifically, the learner consists of four neural networks, i.e., a local policy network, a local critic network, a target policy network and a target critic network, the parameters of which for vehicles are denoted by $\theta_{\mathbf{S}}^{\mu}$, $\theta_{\mathbf{S}}^{Q}$, $\theta_{\mathbf{S}}^{\mu^{\prime}}$, and $\theta_{\mathbf{S}}^{Q^{\prime}}$, respectively.
Similarly, the parameters of the four networks for the edge node are denoted by $\theta_{\mathbf{E}}^{\mu}$, $\theta_{\mathbf{E}}^{Q}$, $\theta_{\mathbf{E}}^{\mu^{\prime}}$ and $\theta_{\mathbf{E}}^{Q^{\prime}}$, respectively.
The parameters of the local policy and local critic networks are randomly initialized.
The parameters of target policy and target critic networks are initialized as the corresponding local network.
$K$ distributed actors are launched to interact with the environment and perform the replay experience storing.
Each actor consists of a local vehicle policy network and a local edge policy network, which are denoted by $\theta_{\mathbf{S}, k}^{\mu}$ and $\theta_{\mathbf{E}, k}^{\mu}$, respectively, and they are replicated from the local policy network of the learner.
The replay buffer $\mathcal{B}$ with a maximum size $|\mathcal{B}|$ is initialized to store replay experiences.

\subsection{Multi-Agent Distributed Policy Execution}

In MAMO, vehicles and the edge node determine their actions via the local policy networks in a distributed way.
The local observation of the system state in vehicle $s$ at time $t$ is represented by 
	\begin{equation}
		\boldsymbol{o}_{s}^{t}=\left\{t, s, l_{s}^t, \mathbf{D}_{s}, \Phi_{s}, \mathbf{D}_{e}^{t}, \mathbf{D}_{\mathbf{V}_e^t}, \boldsymbol{w}^{t}\right\}
	\end{equation} 
\noindent where $t$ is the time slot index; 
$s$ is the vehicle index; $l_{s}^t$ is the location of vehicle $s$; 
$\mathbf{D}_{s}$ represents the set of information that can be sensed by vehicle $s$; 
$\Phi_{s}$ represents the sensing cost of information in $\mathbf{D}_{s}$;
$\mathbf{D}_{e}^{t}$ represents the set of cached information in edge node $e$ at time $t$;
$\mathbf{D}_{\mathbf{V}_e^t}$ represents the set of information required by digital twins modeled in edge node $e$ at time $t$,
and $\boldsymbol{w}^{t}$ represents the weight vector for each objective, which is randomly generated in each iteration.
In particular, $\boldsymbol{w}^{t} = \begin{bmatrix}  w^{(1), t}  &  w^{(2), t} \end{bmatrix}$, where $w^{(1), t} \in (0, 1)$ and $w^{(2), t} \in (0, 1)$ are the weight of the system quality and system profit, respectively, and we have $\sum_{\forall j \in \{1, 2\}} w^{(j), t} = 1$.
On the other hand, the local observation of the system state in edge node $e$ at time $t$ is represented by 
\begin{equation}
	\boldsymbol{o}_{e}^{t}=\left\{t, e, \operatorname{\mathbf{Dis}}_{\mathbf{S}, e}^{t}, \mathbf{D}_{1}, \cdots, \mathbf{D}_{s}, \cdots, \mathbf{D}_{S}, \mathbf{D}_{e}^{t}, \mathbf{D}_{\mathbf{V}_e^t}, \boldsymbol{w}^{t} \right\}
\end{equation}
\noindent where $e$ is the edge node index and $\operatorname{\mathbf{Dis}}_{\mathbf{S}, e}^{t}$ represents the set of distances between vehicles  and edge node $e$.
Therefore, the system state at time $t$ can be represented by $\boldsymbol{o}^{t}=\boldsymbol{o}_{e}^{t} \cup \boldsymbol{o}_{1}^{t} \cup \ldots \cup \boldsymbol{o}_{s}^{t} \cup \ldots \cup \boldsymbol{o}_{S}^{t}$.

The action of vehicle $s$ is represented by 
\begin{equation}
	\boldsymbol{a}_{s}^{t} = \{ \mathbf{C}_s^t,  \{ \lambda_{d, s}^{t}, p_{d, s}^{t} \mid \forall d \in \mathbf{D}_{s}^t \} , \pi_s^t   \}
\end{equation}
where $\mathbf{C}_s^t$ is the sensing  decision; $\lambda_{d, s}^{t}$ and $p_{d, s}^{t}$ are the sensing frequency and uploading priority of information $d \in \mathbf{D}_s^t$, respectively, and $\pi_s^t$ is the transmission power of vehicle $s$ at time $t$.
The actions of vehicles are generated by the local vehicle policy network based on their local observations of the system state.
\begin{equation}
	\boldsymbol{a}_{s}^{t}=\mu_{\mathbf{S}}\left(\boldsymbol{o}_{s}^{t} \mid \theta_{\mathbf{S}}^{\mu}\right)+\epsilon_{s} \mathcal{N}_{s}^{t}
\end{equation}
\noindent where $\mathcal{N}_{s}^{t}$ is an exploration noise to increase the diversity of vehicle actions and $\epsilon_{s}$ is an exploration constant for vehicle $s$.
The set of vehicle actions is denoted by $\boldsymbol{a}_{\mathbf{S}}^{t} = \left\{\boldsymbol{a}_{s}^{t}\mid \forall s \in \mathbf{S}\right\}$.
Then, the action of edge node $e$ is represented by
\begin{equation}
	\boldsymbol{a}_{e}^{t} = \{b_{s, e}^{t} \mid \forall s \in \mathbf{S}_{e}^{t}\}
\end{equation}
where $b_{s, e}^t$ is the V2I bandwidth allocated by edge node $e$ for vehicle $s$ at time $t$.
Similarly, the action of the edge node $e$ can be obtained by the local edge policy network based on the system state as well as the actions of vehicles.
\begin{equation}
	\boldsymbol{a}_{e}^{t}=\mu_{\mathbf{E}}\left(\boldsymbol{o}_{e}^{t},  \boldsymbol{a}_{\boldsymbol{\mathbf{S}}}^{t} \mid \theta_{\mathbf{E}}^{\mu}\right)+\epsilon_{e} \mathcal{N}_{e}^{t}
\end{equation}
\noindent where $\mathcal{N}_{e}^{t}$ and $\epsilon_{e}$ are the exploration noise and exploration constant for the edge node $e$, respectively.
Further, the joint action of vehicles and the edge node is denoted by $\boldsymbol{a}^{t}= \left\{\boldsymbol{a}_{e}^{t}, \boldsymbol{a}_{1}^{t}, \ldots, \boldsymbol{a}_{s}^{t}, \ldots, \boldsymbol{a}_{S}^{t}\right\}$.

The environment obtains the system reward vector by executing the joint action, which is represented by 
	\begin{equation}
	\boldsymbol{r}^{t} = \begin{bmatrix}  r^{(1)}\left(\boldsymbol{a}_{\mathbf{S}}^{t},\boldsymbol{a}_{e}^{t} \mid \boldsymbol{o}^{t}\right)  &  r^{(2)}\left(\boldsymbol{a}_{\mathbf{S}}^{t},\boldsymbol{a}_{e}^{t} \mid \boldsymbol{o}^{t}\right) \end{bmatrix} ^{T}
	\end{equation}
	\noindent where $r^{(1)}\left(\boldsymbol{a}_{\mathbf{S}}^{t},\boldsymbol{a}_{e}^{t} \mid \boldsymbol{o}^{t}\right)$ and $r^{(2)}\left(\boldsymbol{a}_{\mathbf{S}}^{t},\boldsymbol{a}_{e}^{t} \mid \boldsymbol{o}^{t}\right)$ are the reward of two objectives (i.e., the achieved system quality and system profit), respectively, which can be computed by  
	\begin{equation}
		\begin{cases}
			r^{(1)}\left(\boldsymbol{a}_{\mathbf{S}}^{t},\boldsymbol{a}_{e}^{t} \mid \boldsymbol{o}^{t}\right)={1}/{\left|\mathbf{V}_e^t\right|} \sum_{\forall v \in \mathbf{V}_e^t}\operatorname{QDT}_{v}  \\
			r^{(2)}\left(\boldsymbol{a}_{\mathbf{S}}^{t},\boldsymbol{a}_{e}^{t} \mid \boldsymbol{o}^{t}\right)={1}/{\left|\mathbf{V}_e^t\right|} \sum_{\forall v \in \mathbf{V}_e^t} \operatorname{PDT}_{v} 
		\end{cases}
	\end{equation}
Accordingly, the reward of vehicle $s$ in the $j$-th objective is obtained by the difference reward (DR) \cite{foerster2018counterfactual} based reward assignment, which is the difference between the system reward and the reward achieved without its action and can be represented by 
\begin{equation}
\label{equa_dr_sensor_node}
r_{s}^{(j), t}=r^{(j)}\left(\boldsymbol{a}_{\mathbf{S}}^{t},\boldsymbol{a}_{e}^{t} \mid \boldsymbol{o}^{t}\right)-r^{(j)}\left(\boldsymbol{a}_{\mathbf{S}-s}^{t},\boldsymbol{a}_{e}^{t} \mid \boldsymbol{o}^{t}\right), \forall j \in \{1, 2\}
\end{equation}
\noindent where $r^{(j)}\left(\boldsymbol{a}_{\mathbf{S}-s}^{t},\boldsymbol{a}_{e}^{t} \mid \boldsymbol{o}^{t}\right)$ is the system reward achieved without the contribution of vehicle $s$, and it can be obtained by setting null action set for vehicle $s$.
The reward vector of vehicle $s$ at time $t$ is denoted by $\boldsymbol{r}_{s}^{t}$ and  represented by $\boldsymbol{r}_{s}^{t} = \begin{bmatrix}  r_{s}^{(1), t}  &  r_{s}^{(2), t} \end{bmatrix} ^{T}$.
The set of the difference reward for vehicles is denoted by $\boldsymbol{r}_{\mathbf{S}}^{t}=\{ \boldsymbol{r}_{s}^{t} \mid \forall s \in \mathbf{S}\}$.

On the other hand, the system reward is further transformed into a normalized reward for the edge node via min-max normalization.
The reward of edge node $e$ in the $j$-th objective at time $t$ is computed by 
\begin{equation}
	r_{e}^{(j), t}= \frac{r^{(j)}\left(\boldsymbol{a}_{\mathbf{S}}^{t},\boldsymbol{a}_{e}^{t} \mid \boldsymbol{o}^{t}\right) - \min \limits_{\forall {\boldsymbol{a}_{e}^{t}}^{\prime}} r^{(j)}\left(\boldsymbol{a}_{\mathbf{S}}^{t}, {\boldsymbol{a}_{e}^{t}}^{\prime} \mid \boldsymbol{o}^{t}\right)} {\max \limits_{\forall {\boldsymbol{a}_{e}^{t}}^{\prime}} r^{(j)}\left(\boldsymbol{a}_{\mathbf{S}}^{t}, {\boldsymbol{a}_{e}^{t}}^{\prime} \mid \boldsymbol{o}^{t}\right) - \min \limits_{\forall {\boldsymbol{a}_{e}^{t}}^{\prime}} r^{(j)}\left(\boldsymbol{a}_{\mathbf{S}}^{t}, {\boldsymbol{a}_{e}^{t}}^{\prime} \mid \boldsymbol{o}^{t}\right)}
\end{equation}
\noindent where $\min \limits_{\forall {\boldsymbol{a}_{e}^{t}}^{\prime}} r^{(j)} (\boldsymbol{a}_{\mathbf{S}}^{t}, {\boldsymbol{a}_{e}^{t}}^{\prime} \mid \boldsymbol{o}^{t})$ and $\max \limits_{\forall {\boldsymbol{a}_{e}^{t}}^{\prime}} r^{(j)}(\boldsymbol{a}_{\mathbf{S}}^{t}, {\boldsymbol{a}_{e}^{t}}^{\prime} \mid \boldsymbol{o}^{t})$ are the minimum and maximum of the system reward achieved with the unchanged vehicle actions $\boldsymbol{a}_{\mathbf{S}}^{t}$ under the same system state $\boldsymbol{o}^{t}$, respectively.
The reward vector of edge node $e$ at time $t$ is denoted by $\boldsymbol{r}_{e}^{t}$, and represented by $\boldsymbol{r}_{e}^{t} = \begin{bmatrix}  r_{e}^{(1), t}  &  r_{e}^{(2), t} \end{bmatrix} ^{T}$.
The interaction experiences including the system state $\boldsymbol{o}^{t}$, vehicle actions $\boldsymbol{a}_{\mathbf{S}}^{t}$, edge action ${a}_{e}^{t}$, vehicle rewards $\boldsymbol{r}_{S}^{t}$, edge reward $\boldsymbol{r}_{e}^{t}$, weights $\boldsymbol{w}^{t}$, and next system state $\boldsymbol{o}^{t+1}$, are stored into the replay buffer $\mathcal{B}$.
The interaction will continue until the training process of the learner is completed.

\subsection{Multi-Objective Policy Evaluation}

In this section, we propose the dueling critic network (DCN) to evaluate agent action based on the value of state and the advantage of action.
There are two fully-connected networks, namely, the action-advantage (AA) network and the state-value (SV) network in the DCN.
Note that the parameter of AA network for vehicles and the edge node are denoted by $\theta_{\mathbf{S}}^{\mathscr{A}}$ and $\theta_{\mathbf{E}}^{\mathscr{A}}$, respectively.
Similarly, the parameter of the SV network for vehicles and the edge node are denoted by $\theta_{\mathbf{S}}^{\mathscr{V}}$ and $\theta_{\mathbf{E}}^{\mathscr{V}}$, respectively.
We denote the output scalar of AA network with the input of vehicle $s$ by $A_{\mathbf{S}}\left({o}_{s}^{m},  {a}_{s}^{m}, \boldsymbol{a}_{\boldsymbol{\mathbf{S}}-s}^{m}, \boldsymbol{w}^{m} \mid \theta_{\mathbf{S}}^{\mathscr{A}} \right)$, where $\boldsymbol{a}_{\boldsymbol{\mathbf{S}}-s}^{m}$ denotes the action of other vehicles.
Similarly, the output scalar of AA network with the input of edge node $e$ is denoted by $A_{\mathbf{E}}\left({o}_{e}^{m},  {a}_{e}^{m}, \boldsymbol{a}_{\boldsymbol{\mathbf{S}}}^{m}, \boldsymbol{w}^{m} \mid \theta_{\mathbf{E}}^{\mathscr{A}} \right)$, where $\boldsymbol{a}_{\boldsymbol{\mathbf{S}}}^{m}$ denotes the action of all vehicles.
The output scalar of SV network of vehicle $s$ is denoted by $V\left({o}_{s}^{m}, \boldsymbol{w}^{m} \mid \theta_{\mathbf{S}}^{\mathscr{V}} \right)$. 
Similarly, the output scalar of SV network of edge node $e$ is denoted by $V\left({o}_{e}^{m}, \boldsymbol{w}^{m} \mid \theta_{\mathbf{E}}^{\mathscr{V}} \right)$.

The agent action evaluation consists of three steps.
First, the AA network estimates the advantage function by outputting the advantage of agent action based on the observation, action, and weights.
Second, the VS network estimates the value function by outputting the value of state according to the observation and weights.
Third, an aggregating module is adopted to output a single value to evaluate the action based on the advantage of action and the value of state.
Specifically, $N$ actions are randomly generated and replaced with the agent action in the AA network to evaluate the average action value of random actions.
We denote the $n$-th random action of vehicle $s$ and edge node $e$ by ${a}_{s}^{m, n}$ and ${a}_{e}^{m, n}$, respectively.
Therefore, the advantage of the $n$-th random action of vehicle $s$ and edge node $e$ can be represented by $A_{\mathbf{S}}\left({o}_{s}^{m},  {a}_{s}^{m, n}, \boldsymbol{a}_{\boldsymbol{\mathbf{S}}-s}^{m}, \boldsymbol{w}^{m} \mid \theta_{S}^{\mathscr{A}} \right)$ and $A_{\mathbf{E}}\left({o}_{e}^{m},  {a}_{e}^{m, n}, \boldsymbol{a}_{\boldsymbol{\mathbf{S}}}^{m}, \boldsymbol{w}^{m} \mid \theta_{\mathbf{E}}^{\mathscr{A}} \right)$, respectively.

The aggregating module of the Q function is constructed by evaluating the advantage of the agent action over the average advantage of random actions.
Thus, the action values of vehicle $s \in \mathbf{S}$ and edge node $e$ are computed by 
\begin{equation}
    \begin{aligned}
       &Q_{\mathbf{S}}\left({o}_{s}^{m}, {a}_{s}^{m}, \boldsymbol{a}_{\boldsymbol{\mathbf{S}}-s}^{m}, \boldsymbol{w}^{m} \mid \theta_{\mathbf{S}}^{Q} \right)\\
       &= V\left({o}_{s}^{m}, \boldsymbol{w}^{m} \mid \theta_{\mathbf{S}}^{\mathscr{V}} \right) + A_{\mathbf{S}}\left({o}_{s}^{m},  {a}_{s}^{m}, \boldsymbol{a}_{\boldsymbol{\mathbf{S}}-s}^{m}, \boldsymbol{w}^{m} \mid \theta_{\mathbf{S}}^{\mathscr{A}} \right)\\
        &- \frac{1}{N} \sum_{\forall n} A_{\mathbf{S}}\left({o}_{s}^{m},  {a}_{s}^{m, n}, \boldsymbol{a}_{\boldsymbol{\mathbf{S}}-s}^{m}, \boldsymbol{w}^{m} \mid \theta_{\mathbf{S}}^{\mathscr{A}} \right)
    \end{aligned}
\end{equation}
\begin{equation}
    \begin{aligned}
       &Q_{E}\left({o}_{e}^{m},  {a}_{e}^{m}, \boldsymbol{a}_{\boldsymbol{\mathbf{S}}}^{m}, \boldsymbol{w}^{m} \mid \theta_{\mathbf{E}}^{Q} \right)\\
       &= V\left({o}_{e}^{m}, \boldsymbol{w}^{m} \mid \theta_{\mathbf{E}}^{\mathscr{V}} \right) + A_{\mathbf{E}}\left({o}_{e}^{m},  {a}_{e}^{m}, \boldsymbol{a}_{\boldsymbol{\mathbf{S}}}^{m}, \boldsymbol{w}^{m} \mid \theta_{\mathbf{E}}^{\mathscr{A}} \right)\\
        &- \frac{1}{N} \sum_{\forall n} A_{\mathbf{E}}\left({o}_{e}^{m},  {a}_{e}^{m, n}, \boldsymbol{a}_{\boldsymbol{\mathbf{S}}}^{m}, \boldsymbol{w}^{m} \mid \theta_{\mathbf{E}}^{\mathscr{A}} \right)
    \end{aligned}
\end{equation}
where $\theta_{\mathbf{S}}^{Q}$ and $\theta_{\mathbf{S}}^{Q}$ contain the parameters of the corresponding AA and SV networks. 
\begin{equation}
	\begin{aligned}
		\theta_{\mathbf{S}}^{Q} = (\theta_{\mathbf{S}}^{\mathscr{A}}, \theta_{\mathbf{S}}^{\mathscr{V}}), \theta_{\mathbf{S}}^{Q^{\prime}} = (\theta_{\mathbf{S}}^{\mathscr{A}^{\prime}}, \theta_{\mathbf{S}}^{\mathscr{V}^{\prime}}) \\
		\theta_{\mathbf{E}}^{Q} = (\theta_{\mathbf{E}}^{\mathscr{A}}, \theta_{\mathbf{E}}^{\mathscr{V}}), \theta_{\mathbf{E}}^{Q^{\prime}} = (\theta_{\mathbf{E}}^{\mathscr{A}^{\prime}}, \theta_{\mathbf{E}}^{\mathscr{V}^{\prime}})
	\end{aligned}
\end{equation}

\subsection{Network Learning and Updating}

A minibatch of $M$ transitions is sampled from the replay buffer $\mathcal{B}$ to train the policy and critic networks of vehicles and the edge node, which is denoted by $\left(\boldsymbol{o}_{\mathbf{S}}^{m}, {o}_{e}^{m}, \boldsymbol{w}^{m}, \boldsymbol{a}_{\mathbf{S}}^{m}, {a}_{e}^{m}, \boldsymbol{r}_{\mathbf{S}}^{m}, \boldsymbol{r}_{e}^{m}, \boldsymbol{o}_{\mathbf{S}}^{m+1}, {o}_{e}^{m+1}, \boldsymbol{w}^{m+1}\right)$.
The target value of vehicle $s$ is denoted by
\begin{equation}
	y_{s}^{m} = \boldsymbol{r}_{s}^{m} \boldsymbol{w}^{m} +\gamma Q_{\mathbf{S}}^{\prime}\left({o}_{s}^{m+1},  {a}_{s}^{m+1}, \boldsymbol{a}_{\boldsymbol{\mathbf{S}}-s}^{m+1}, \boldsymbol{w}^{m+1} \mid \theta_{\mathbf{S}}^{Q^{\prime}} \right)
\end{equation}
\noindent where $Q_{\mathbf{S}}^{\prime}({o}_{s}^{m+1},  {a}_{s}^{m+1}, \boldsymbol{a}_{\boldsymbol{\mathbf{S}}-s}^{m+1}, \boldsymbol{w}^{m+1} \mid \theta_{\mathbf{S}}^{Q^{\prime}})$ is the action value generated by the target vehicle critic network;
$\gamma$ is the discount;
$\boldsymbol{a}_{\boldsymbol{\mathbf{S}}-s}^{m+1}$ is the next vehicle actions without the vehicle $s$, i.e., $\boldsymbol{a}_{\boldsymbol{\mathbf{S}}-s}^{m+1} = \{ {a}_{1}^{m+1}, \ldots, {a}_{s-1}^{m+1}, {a}_{s+1}^{m+1}, \ldots, {a}_{S}^{m+1} \}$, and ${a}_{s}^{m+1}$ is the next action of vehicle $s$ generated by the target vehicle policy network based on the local observation of next system state, i.e., ${a}_{s}^{m+1} = \mu_{\mathbf{S}}^{\prime}(\boldsymbol{o}_{s}^{m+1} \mid \theta_{\mathbf{S}}^{\mu^{\prime}})$.
Similarly, the target value of edge node $e$ is denoted by
\begin{equation}
	y_{e}^{m} = \boldsymbol{r}_{e}^{m} \boldsymbol{w}^{m} +\gamma Q_{\mathbf{E}}^{\prime}\left({o}_{e}^{m+1},  {a}_{e}^{m+1}, \boldsymbol{a}_{\boldsymbol{\mathbf{S}}}^{m+1}, \boldsymbol{w}^{m+1} \mid \theta_{\mathbf{E}}^{Q^{\prime}} \right)
\end{equation}
\noindent where $Q_{\mathbf{E}}^{\prime}({o}_{e}^{m+1},  {a}_{e}^{m+1}, \boldsymbol{a}_{\boldsymbol{\mathbf{S}}}^{m+1}, \boldsymbol{w}^{m+1} \mid \theta_{\mathbf{E}}^{Q^{\prime}})$ denotes the action value generated by the target edge critic network; $\boldsymbol{a}_{\boldsymbol{\mathbf{S}}}^{m+1}$ is the next vehicle actions,
and ${a}_{e}^{m+1}$ denotes the next edge action, which can be obtained by the target edge policy network based on its local observation of the next system state, i.e., ${a}_{e}^{m+1} = \mu_{\mathbf{E}}^{\prime}(\boldsymbol{o}_{e}^{m+1}, \boldsymbol{a}_{\mathbf{S}}^{m+1} \mid \theta_{\mathbf{E}}^{\mu^{\prime}})$.

The loss function of the vehicle critic network and edge critic network are obtained by the categorical distribution temporal difference (TD) learning, which is represented by 
\begin{equation}
	\mathcal{L}\left(\theta_{\mathbf{S}}^{Q}\right)=\frac{1}{M} \sum_{m} \frac{1}{S} \sum_{s} {Y_s^{m}}
\end{equation}
\begin{equation}
	\mathcal{L}\left(\theta_{\mathbf{E}}^{Q}\right)=\frac{1}{M} \sum_{m} {Y_e^{m}}
\end{equation}
\noindent where $Y_s^{m}$ and $Y_e^{m}$ are the squares of the difference between the target value and the action value generated by the local critic network for vehicle $s$ and edge node $e$, respectively. 
\begin{equation}
	\begin{aligned}
		Y_s^{m} &= \left(y_{s}^{m}-Q_{\mathbf{S}}\left({o}_{s}^{m},  {a}_{s}^{m}, \boldsymbol{a}_{\boldsymbol{\mathbf{S}}-s}^{m}, \boldsymbol{w}^{m} \mid \theta_{\mathbf{S}}^{Q} \right)\right)^{2} \\
	\end{aligned}
\end{equation}
\begin{equation}
	\begin{aligned}
		Y_e^{m} &=\left(y_{e}^{m}-Q_{\mathbf{E}}\left({o}_{e}^{m},  {a}_{e}^{m}, \boldsymbol{a}_{\boldsymbol{\mathbf{S}}}^{m}, \boldsymbol{w}^{m} \mid \theta_{\mathbf{S}}^{Q} \right)\right)^{2} \\
	\end{aligned}
\end{equation}
The vehicle and edge policy network parameters are updated via deterministic policy gradient.
\begin{equation}
	\nabla_{\theta_{\mathbf{S}}^{\mu}} \mathcal{J} (\theta_{\mathbf{S}}^{\mu}) \approx \frac{1}{M} \sum_{m} \frac{1}{S} \sum_{s} P_{s}^{m} 
\end{equation}
\begin{equation}
	\nabla_{\theta_{\mathbf{E}}^{\mu}} \mathcal{J} (\theta_{\mathbf{E}}^{\mu}) \approx \frac{1}{M} \sum_{m} P_{e}^{m} 
\end{equation}
\noindent where 
\begin{equation}
\text{\footnotesize$P_{s}^{m} = \nabla_{{a}_{s}^{m}} Q_{\mathbf{S}}\left({o}_{s}^{m}, {a}_{s}^{m}, \boldsymbol{a}_{\boldsymbol{\mathbf{S}}-s}^{m}, \boldsymbol{w}^{m} \mid \theta_{S}^{Q} \right) \nabla_{\theta_{\mathbf{S}}^{\mu}} \mu_{\mathbf{S}}\left({o}_{s}^{m} \mid \theta_{\mathbf{S}}^{\mu}\right)$}
\end{equation}
\begin{equation}
\text{\footnotesize$P_{e}^{m} = \nabla_{{a}_{e}^{m}} Q_{\mathbf{E}}\left({o}_{e}^{m}, {a}_{e}^{m}, \boldsymbol{a}_{\boldsymbol{\mathbf{S}}}^{m}, \boldsymbol{w}^{m} \mid \theta_{\mathbf{E}}^{Q} \right) \nabla_{\theta_{\mathbf{E}}^{\mu}} \mu_{\mathbf{E}}\left({o}_{e}^{m}, {\boldsymbol{a}}_{\boldsymbol{\mathbf{S}}}^{m} \mid \theta_{\mathbf{E}}^{\mu}\right)$}
\end{equation}

The local policy and critic network parameters are updated with the learning rate $\alpha$ and $\beta$, respectively.
In particular, vehicles and the edge node update the parameters of target networks periodically, i.e., when $t \operatorname{mod} t_{\operatorname{tgt}} = 0$, where $t_{\operatorname{tgt}}$ is the parameter updating period of the target networks,
\begin{equation}
	\begin{aligned}
	\text{\footnotesize$\theta_{\mathbf{S}}^{\mu^{\prime}} \leftarrow n_{\mathbf{S}} \theta_{\mathbf{S}}^{\mu}+(1-n_{\mathbf{S}}) \theta_{\mathbf{S}}^{\mu^{\prime}}, \theta_{\mathbf{S}}^{Q^{\prime}} \leftarrow n_{\mathbf{S}} \theta_{\mathbf{S}}^{Q}+(1-n_{\mathbf{S}}) \theta_{\mathbf{S}}^{Q^{\prime}}$}\\
	\text{\footnotesize$\theta_{\mathbf{E}}^{\mu^{\prime}} \leftarrow n_{\mathbf{E}} \theta_{\mathbf{E}}^{\mu}+(1-n_{\mathbf{E}}) \theta_{\mathbf{E}}^{\mu^{\prime}}, \theta_{\mathbf{E}}^{Q^{\prime}} \leftarrow n_{\mathbf{E}} \theta_{\mathbf{E}}^{Q}+(1-n_{\mathbf{E}})  \theta_{\mathbf{E}}^{Q^{\prime}}$}
	\end{aligned}
\end{equation}
\noindent with $n_{\mathbf{S}} \ll 1$ and $n_{\mathbf{E}} \ll 1$.
Similarly, the parameters of policy networks at the distributed actors are  updated periodically, i.e., when $t \operatorname{mod} t_{\operatorname{act}} = 0$, where $t_{\operatorname{act}}$ is the parameter updating period of the policy network at the distributed actors.
\begin{equation}
	\begin{aligned}
		\theta_{\mathbf{S}, k}^{\mu} \leftarrow \theta^{{\mu}^{\prime}}_{\mathbf{S}}, \theta_{\mathbf{S}, k}^{Q} \leftarrow \theta_{\mathbf{S}}^{Q^{\prime}}, \forall k \in \{1, 2, \cdots, K\}\\
		\theta_{\mathbf{E}, k}^{\mu} \leftarrow \theta_{\mathbf{E}}^{\mu^{\prime}}, \theta_{\mathbf{E}, k}^{Q} \leftarrow \theta_{\mathbf{E}}^{Q^{\prime}}, \forall k \in \{1, 2, \cdots, K\}
	\end{aligned}
\end{equation} 

\section{Performance Evaluation} \label{evaluation}

\subsection{Experiment Setup}

In this section, we use Python 3.9.13 and TensorFlow 2.8.0 to evaluate the performance of the proposed MAMO scheme using a Ubuntu 20.04 server with an AMD Ryzen 9 5950X 16-core processor @ 3.4 GHz, two NVIDIA GeForce RTX 3090 GPUs, and 64 GB memory.
We consider the general scenario in a 1$\times$1 $\operatorname{km}^2$ square area, where the realistic vehicular trajectories are utilized as traffic inputs collected from Didi GAIA open data set \cite{didi}.
On the basis of referring to \cite{sadek2009distributed} and \cite{wang2019delay}, the simulation parameter settings are as follows. 
The V2I communication range is set as 500 m.
The transmission power is set as 100 mW.
The AWGN and reliability threshold are set as -90 dBm and 0.9, respectively.
The channel fading gains of V2I communications follow the Gaussian distribution with a mean of 2 and a variance of 0.4
The weighting factors for $\hat{\Theta_{v}}$, $\hat{\Psi_{v}}$, $\hat{\Xi_{v}}$, $\hat{\Phi_{v}}$, and $\hat{\Omega_{v}}$ are set as 0.6, 0.4, 0.2, 0.4, and 0.4, respectively.

For the implementation of the proposed solution, the architectures and hyperparameters of the policy and critic networks are described as follows.
The local policy network is a four-layer fully connected neural network with two hidden layers, where the numbers of neurons are 256 and 128, respectively.
The architecture of the target policy network is the same as the local policy network.
The local critic network is a four-layer fully connected neural network with two hidden layers, where the numbers of neurons are 512 and 256, respectively.
The architecture of the target critic network is the same as the local critic network.
The discount, batch size, and maximum replay buffer size are set to 0.996, 256, and 1$\times10^{6}$, respectively.
The learning rates for policy and critic networks are set to 1$\times10^{-4}$ and 1$\times10^{-4}$, respectively.

Three comparable algorithms are implemented as follows.
\begin{itemize}
	\item RA: it randomly selects one action to determine the sensing information, sensing frequencies, uploading priorities, transmission power, and V2I bandwidth allocation.
	\item D4PG\cite{barth2018distributed}: it implements an agent at the edge node to determine the sensing information, sensing frequencies, uploading priorities, transmission power, and V2I bandwidth allocation in a centralized way based on the system state. The system quality and the system profit weights are set as 0.5 and 0.5, respectively. 
	\item MAD4PG \cite{xu2022joint}: it is a multi-agent version of D4PG, which is implemented in vehicles to decide the sensing information, sensing frequencies, uploading priorities, and transmission power based on local observation of the physical environment, and the edge node to determine the V2I bandwidth allocation. The system quality and the system profit weights are set as 0.5 and 0.5, respectively.
\end{itemize}

To capture the performance of algorithm in terms of the digital twin modeling quality and effectiveness, we design two new metrics as follows.
\begin{itemize}
	\item Quality per unit cost (QPUC): it is defined as the achieved system quality with the unit system cost, which is computed by
		\begin{equation}
			\operatorname{QPUC}=\frac{\sum_{\forall t \in \mathbf{T}} \sum_{\forall e \in \mathbf{E}} \sum_{\forall v \in \mathbf{V}_e^t} \mathrm{QDT}_v}{\sum_{\forall t \in \mathbf{T}} \sum_{\forall e \in \mathbf{E}} \sum_{\forall v \in \mathbf{V}_e^t} \mathrm{CDT}_v}
		\end{equation}
		where $\mathrm{QDT}_v$ and $\mathrm{CDT}_v$ are the quality and cost of digital twin $v$, respectively.
	\item Profit per unit quality (PPUQ): it is defined as the achieved system profit with the unit system quality, which is computed by
		\begin{equation}
		\operatorname{PPUQ}=\frac{\sum_{\forall t \in \mathbf{T}} \sum_{\forall e \in \mathbf{E}} \sum_{\forall v \in \mathbf{V}_e^t}\mathrm{PDT}_v}{\sum_{\forall t \in \mathbf{T}} \sum_{\forall e \in \mathbf{E}} \sum_{\forall v \in \mathbf{V}_e^t} \mathrm{QDT}_v}
		\end{equation}
		where $\mathrm{PDT}_v$ and $\mathrm{CDT}_v$ are the profit and cost of digital twin $v$, respectively.
\end{itemize}
The higher of QPUC indicates it can achieve higher system quality with the same cost, and the higher of PPUQ indicates it can use the sensing and communication resources more efficiently. These two metrics provide a comprehensive indication of the performance of the algorithm in maximizing system quality and minimizing system cost at the same time.
We further design the four metrics named \textit{Average Timeliness} (AT),  \textit{Average Redundancy} (AR), \textit{Average Sensing Cost} (ASC), and \textit{Average Transmission Cost} (ATC) based on Eqs. \ref{definition of timeliness}, \ref{definition of redundancy}, \ref{definition of sensing cost}, and \ref{definition of transmission cost}, respectively. 

\subsection{Result Analysis}

\begin{figure}[t]
 \centering
 \includegraphics[width=1\columnwidth]{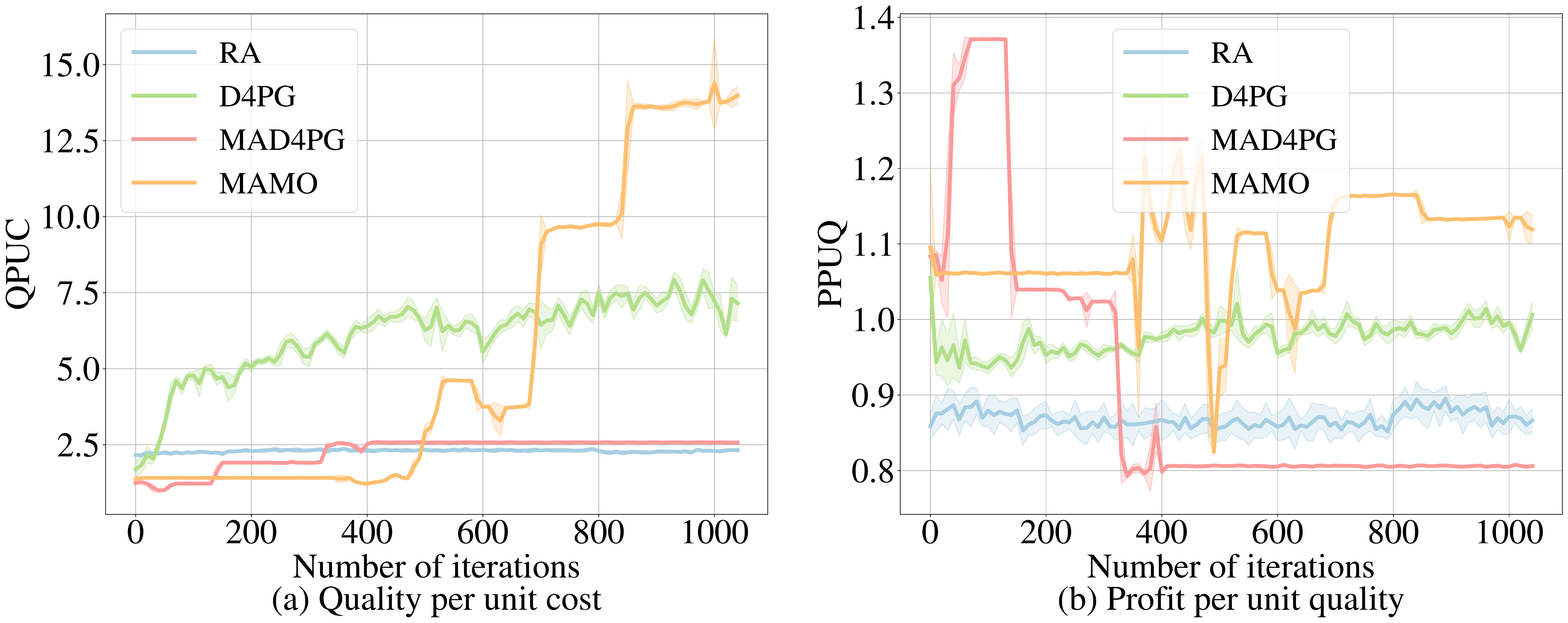}
 \caption{Convergence comparison, which shows MAMO achieves the highest QPUC and the highest PPUQ compared with RA, D4PG, and MAD4PG after convergence (around 850 iterations)}
 \label{fig_algorithms}
\end{figure}

\textit{1) Algorithm convergence:} {Fig. \ref{fig_algorithms} compares the convergence of the four algorithms. In particular, Figs. \ref{fig_algorithms}(a) and \ref{fig_algorithms}(b) compare the QPUC and PPUQ of the four algorithms, respectively, where the X-axis indicates the number of iterations, and the Y-axis indicates the achieved QPUC and PPUQ, respectively. The higher of QPUC and PPUQ mean the better performance on system quality and system cost, respectively. As noted, the proposed MAMO solution achieves the highest QPUC (around 13.6) and the highest PPUQ (around 1.13) after around 850 iterations. In contrast, RA, D4PG and MAD4PG achieve the QPUC around 2.29, 7.34 and 2.58, respectively, and they achieve the PPUQ around 0.87, 0.99 and 0.81, respectively. MAMO achieves about 494.1\%, 85.5\% and 428.8\% improvement with respect to QPUC and about 30.6\%, 14.2\% and 40.7\% improvement with respect to PPUQ compared with RA, D4PG and MAD4PG, respectively.
It is noted that MAMO is the only one solution, which can improve the QPUC and PPUQ at the same time. 
It shows the advantage of MAMO in terms of maximizing the QPUC and PPUQ simultaneously.

\begin{figure}[htbp]
 \centering
 \includegraphics[width=1\columnwidth]{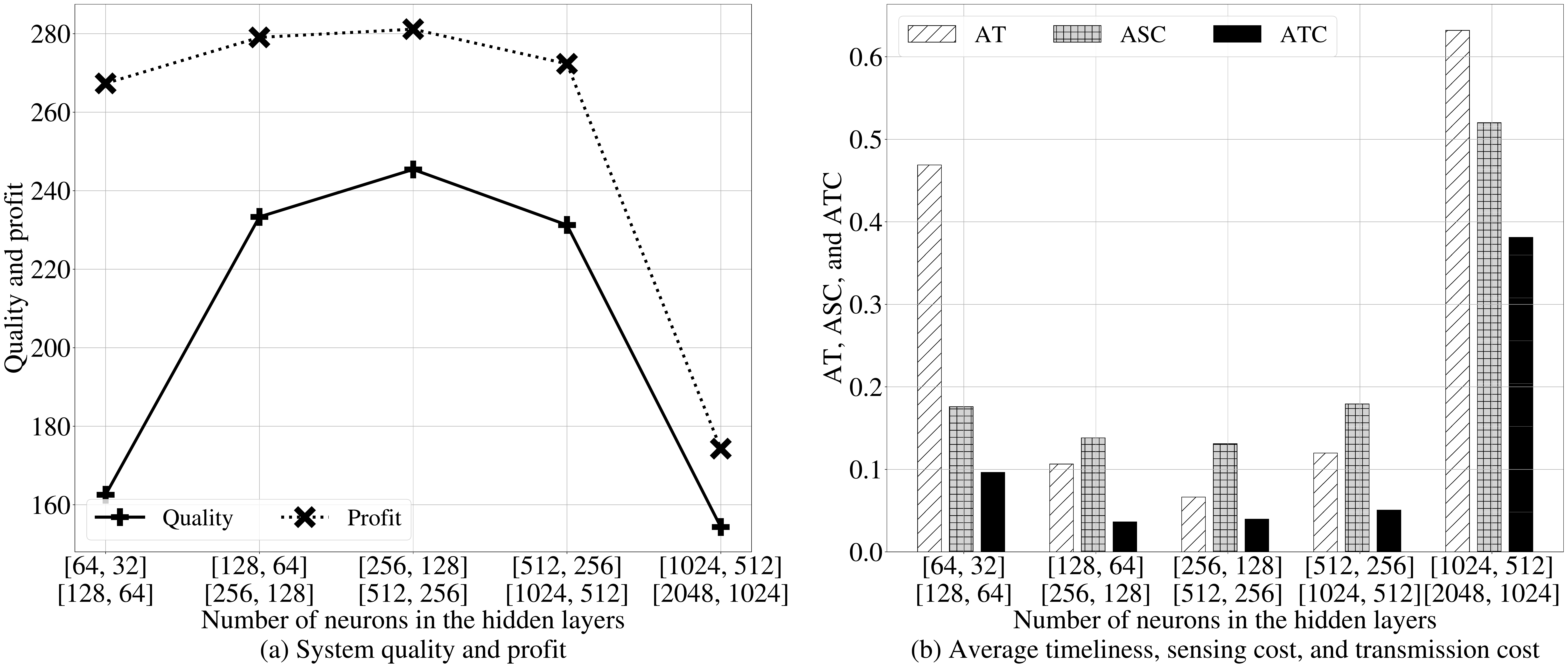}
 \caption{Performance comparison of MAMO under different numbers of neurons in the hidden layers}
 \label{fig_networks}
\end{figure}

\textit{2) Effect of number of neurons:}
Fig. \ref{fig_networks} compares the performance of MAMO under different numbers of neurons, where the X-axis indicates the number of neurons in the two hidden layers of the policy network and critic network, which are set to [64, 32] $\sim$ [1024, 512], and [128, 64] $\sim$ [2048, 1024], respectively. As demonstrated in Fig. \ref{fig_networks}(a), MAMO achieves the highest system quality and the highest system profit when the number of neurons in the hidden layers of the policy and critic networks are set as the default settings (i.e., [256, 128] and [512, 256]). Fig. \ref{fig_networks}(b) compares the other three metrics including AT, ASC and ATC. The lower of AT, ASC and ATC mean the better performance on information freshness, sensing cost and transmission cost, respectively. It is noted that MAMO achieves the best performance in terms of minimizing the AT, ASC, and ATC with the number of neurons in each hidden layer set in the default setting.

\begin{figure}[htbp]
 \centering
 \includegraphics[width=1\columnwidth]{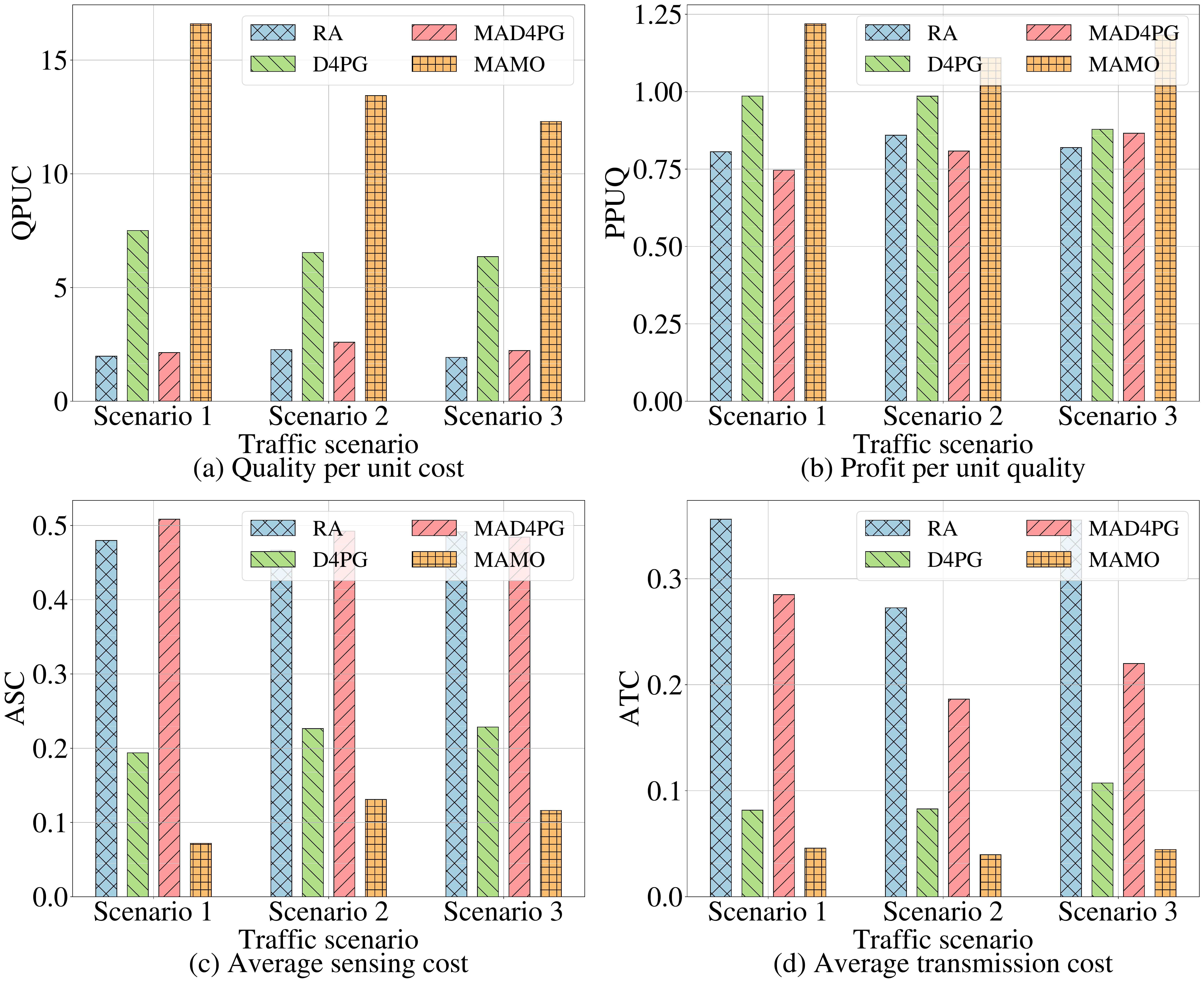}
 \caption{Performance comparison under different traffic scenarios}
 \label{fig_scenarios}
\end{figure}

\textit{3) Effect of traffic scenarios:} 
Fig. \ref{fig_scenarios} compares the performance of the four algorithms under different traffic scenarios, where the X-axis indicates the traffic scenario, in which the realistic vehicular trajectories are extracted from different time and spaces, i.e., 1): a 1 km $\times$ 1 km area of Qingyang District, Chengdu, China, from 8:00 to 8:05, on 16 Nov. 2016; 2): the same area from 23:00 to 23:05, on 16 Nov. 2016; 3): a 1 km $\times$ 1 km area of Beilin District, Xian, China, from 8:00 to 8:05, on 27 Nov. 2016. Fig. \ref{fig_scenarios}(a) compares the QPUC of the four algorithms. As demonstrated, MAMO achieves the highest QPUC under all scenarios. Fig. 5(b) compares the PPUQ of the four algorithms. It is expected that MAMO achieves the highest PPUQ under all the scenarios. In particular, the proposed MAMO solution improves the QPUC by around 589.0\%, 106.7\% and 514.8\% and improves the PPUQ by around 41.6\%, 23.6\% and 45.7\% over RA, D4PG and MAD4PG, respectively.
Fig \ref{fig_scenarios}(c) compares the ASC of the four algorithms.
As noted, the ASC of MAMO is lower than RA, D4PG and MAD4PG.
It demonstrates that the MAMO can cooperate among vehicles by sensing information cooperatively.
Fig. \ref{fig_scenarios}(d) compares the ATC of the four algorithms.
As noted, the ATC of MAMO is the lowest under different scenarios.

\begin{figure}[htbp]
 \centering
 \includegraphics[width=1\columnwidth]{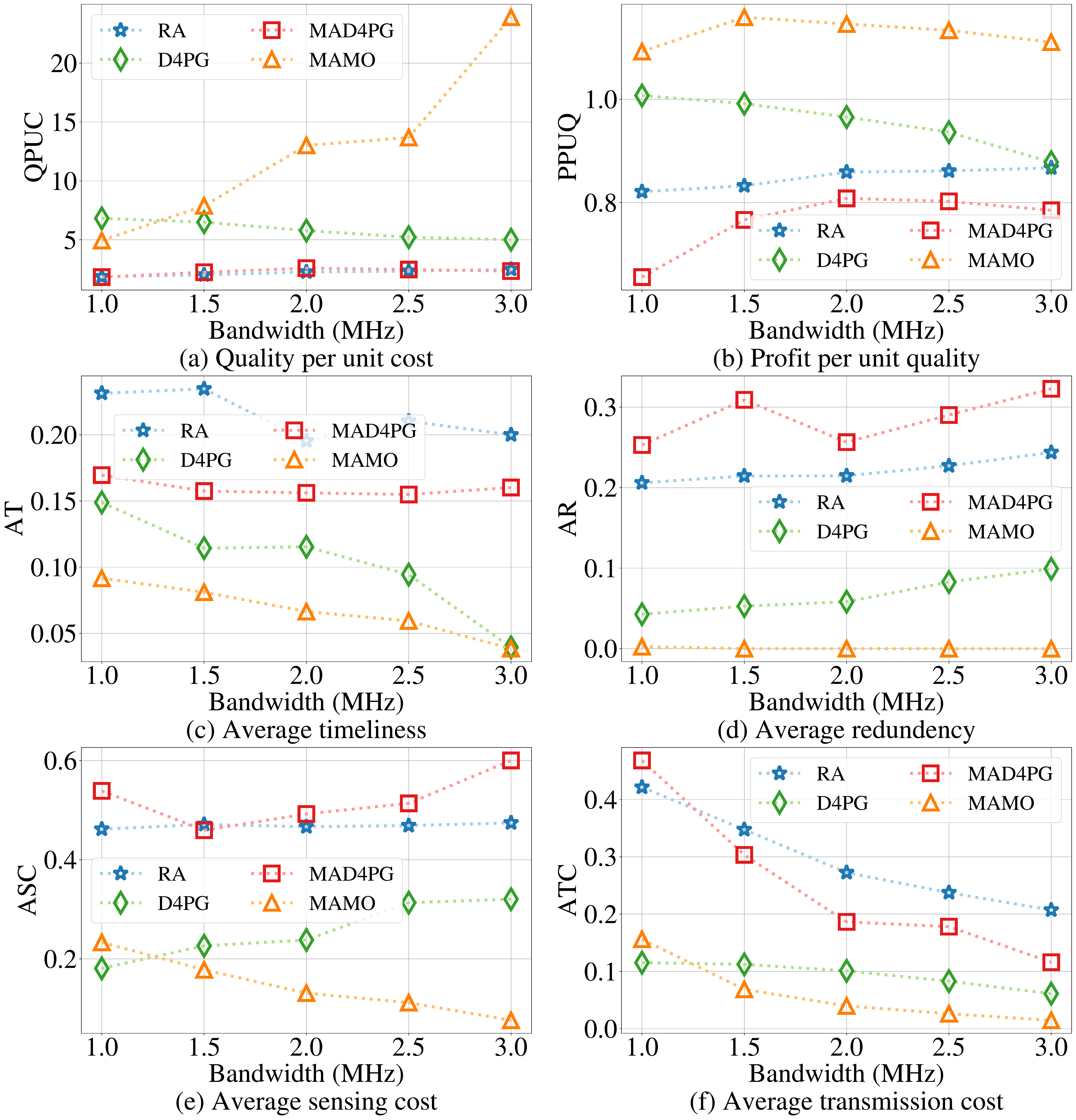}
 \caption{Performance comparison under different V2I bandwidths}
 \label{fig_bandwidth}
\end{figure}

\textit{4) Effect of V2I bandwidths:} 
Fig. \ref{fig_bandwidth} compares the performance of the four algorithms under different V2I bandwidths, where the X-axis indicates the V2I bandwidth, which increases from 1 MHz to 3 MHz. A larger V2I bandwidth represents that the allocated V2I bandwidth for each vehicle can be enlarged. Fig. \ref{fig_bandwidth}(a) compares the QPUC of the four algorithms. As the bandwidth increases, the QPUC of MAMO increases accordingly. It is because the cooperation of sensing and uploading among vehicles is more efficient in MAMO with rich bandwidth. The advantage can be further justified by Fig. \ref{fig_bandwidth}(b), which shows the PPUQ of the four algorithms. As demonstrated, MAMO achieves the highest PPUQ under different V2I bandwidths. In particular, MAMO improves the QPUC by around 453.3\%, 131.4\% and 437.6\% and improves the PPUQ by around 33.0\%, 18.3\% and 48.4\% over RA, D4PG and MAD4PG, respectively.
Fig. \ref{fig_bandwidth}(c) compares the AT of the four algorithms.
It is expected that MAMO achieves the lowest AT. 
It is observed that the performance difference between MAMO and D4PG is small when the bandwidth increases from 2.5 MHz to 3 MHz.
This is because the improvement of timeliness of digital twins with enrich bandwidth is limited.
Fig. \ref{fig_bandwidth}(d) compares the AR of the four algorithms.
The lower of AR means the better performance on cooperative sensing and uploading, and MAMO is expected to achieve the lowest AR.
Figs. \ref{fig_bandwidth}(e) and \ref{fig_bandwidth}(f) compare the ASC and ATC of the four algorithms, respectively.
It is observed that the ATC of the four algorithms decreases when the bandwidth increases.
The reason is that when the bandwidth increases, the information uploading time decreases, resulting in a lower transmission cost.
As expected, the ASC and ATC of MAMO remain at the lowest level in most cases.

\textit{5) Effect of digital twin requirements:}
Fig. \ref{fig_information_number} compares the performance of the four algorithms under different digital twin requirements, where the X-axis indicates the average number of information required by digital twins, which increases from 3 to 7. A larger average number of the required information of digital twin indicates that the vehicles have more significant sensing and uploading workloads. Fig. \ref{fig_information_number}(a) compares the QPUC of the four algorithms. With the increasing average required information number, the QPUC of the four algorithms decreases accordingly. As noted, MAMO keeps the highest QPUC in all cases. Fig. \ref{fig_information_number}(b) compares the PPUQ of the four algorithms. As expected, MAMO achieves the highest PPUQ across all cases. In particular, the QPUC of MAMO outperforms RA, D4PG and MAD4PG by around 458.7\%, 130.6\% and 426.2\%, respectively, and the PPUQ of MAMO outperforms RA, D4PG and MAD4PG by around 31.5\%, 18.2\% and 40.7\%, respectively.
Fig. \ref{fig_information_number}(c) compares the AT of the four algorithms.
As expected, MAMO achieves the best performance in terms of AT.
Fig. \ref{fig_information_number}(d) compares the AR of the four algorithms, which shows that MAMO can achieve the lowest AR across all cases.
Figs. \ref{fig_information_number}(e) and \ref{fig_information_number}(f) compare the ASC and ATC of the four algorithms, respectively.
It is noted that the ASC and ATC of the four algorithms increase when the average information number increases.
The reason is that the average amount of information required by DT-VEC increases, resulting in higher vehicle sensing and transmission costs.

\begin{figure}[htbp]
 \centering
 \includegraphics[width=1\columnwidth]{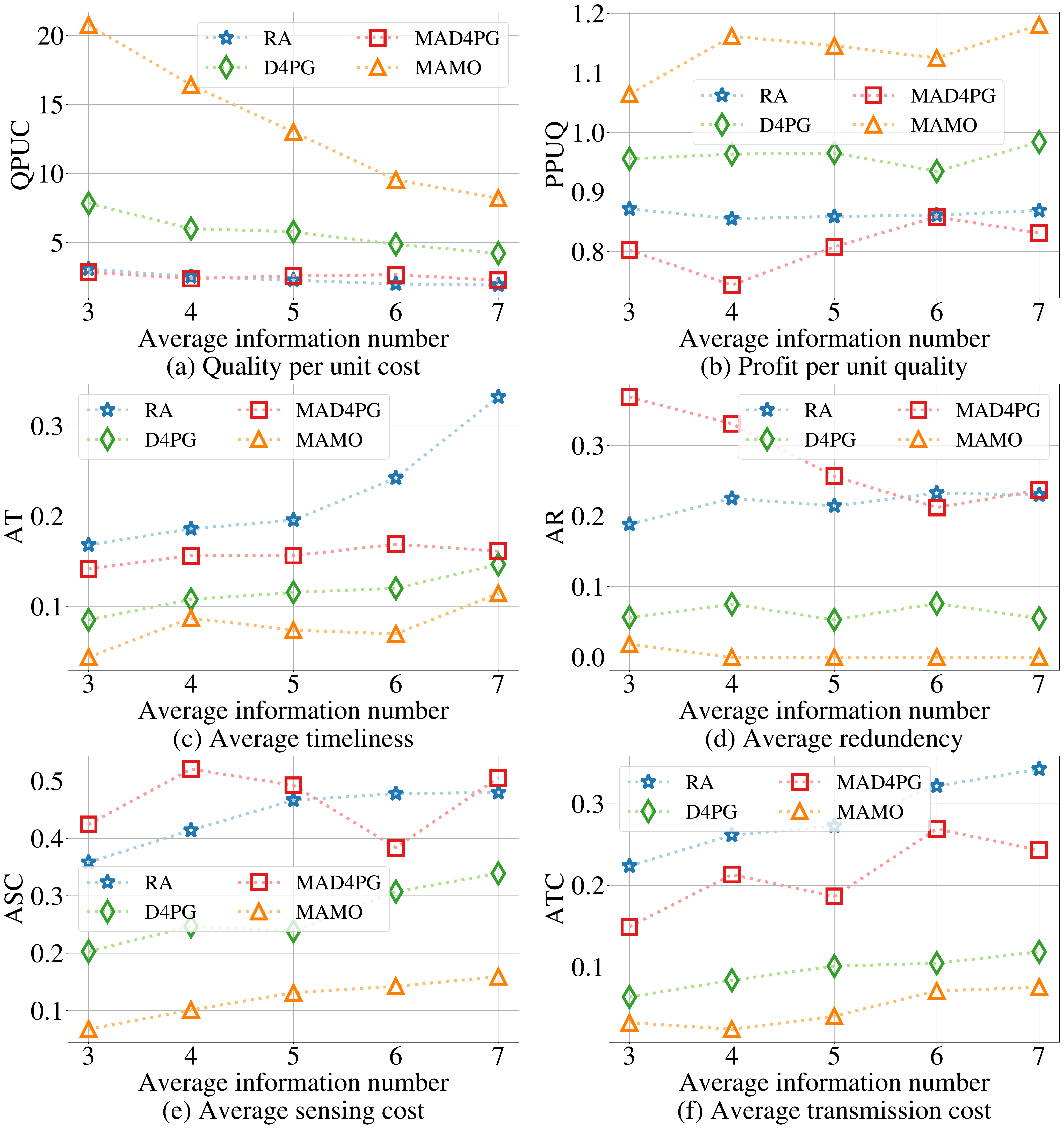}
 \caption{Performance comparison under different digital twin requirements}
 \label{fig_information_number}
\end{figure}

\section{Conclusion and Future Work} \label{conclusion}
In this paper, the DT-VEC architecture was presented, in which heterogeneous information is sensed and uploaded via V2I communications.
The digital twins are modeled based on information fusion, which are further utilized to form logical views in edge nodes and reflect the physical vehicular environment.
The cooperative sensing model was derived based on multi-class M/G/1 priority queue, and the V2I uploading model was derived based on channel fading distribution and SNR threshold.
On this basis, two metrics, QDT and CDT, were designed to measure the quality and cost of digital twins modeled at the edge nodes, and a bi-objective problem was formulated to maximize the system quality and minimize the system cost via cooperative sensing and uploading.
Further, a solution based on multi-agent multi-objective deep reinforcement learning was proposed, where a dueling critic network was adopted to to evaluate agent action based on the value of state and the advantage of action. 
Finally, a comprehensive performance evaluation was conducted, which demonstrated the superiority of the proposed solution.

In the future work, the cooperation among the edge nodes will be further investigated to enhance overall system performance. In addition, we would like to implement the solution model in the real world to verify system capability based on the actual vehicular network environments.

%

\bibliographystyle{IEEEtran} 
\bibliography{reference}

\begin{IEEEbiography}[{\includegraphics[width=1in,height=1.25in,clip,keepaspectratio]{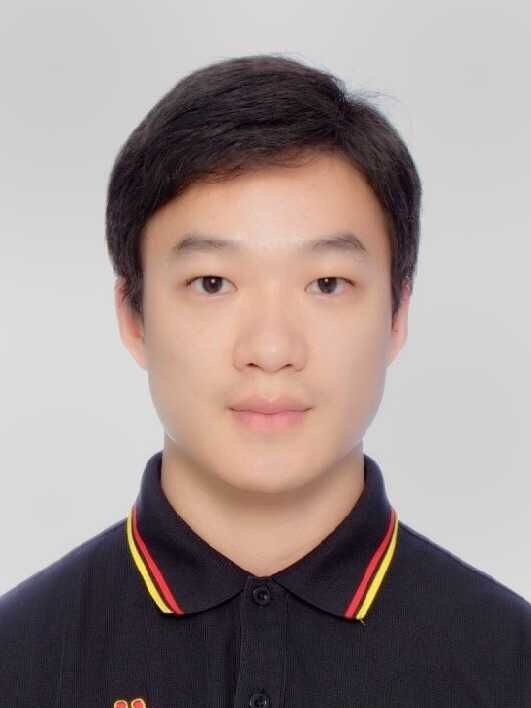}}]{Kai Liu}
(Senior Member, IEEE) received the Ph.D. degree in computer science from the City University of Hong Kong in 2011. He is currently a Full Professor with the College of Computer Science, Chongqing University, China. From 2010 to 2011, he was a Visiting Scholar with the Department of Computer Science, University of Virginia, Charlottesville, VA, USA. From 2011 to 2014, he was a Postdoctoral Fellow with Nanyang Technological University, Singapore, City University of Hong Kong, and Hong Kong Baptist University, Hong Kong. His research interests include mobile computing, pervasive computing, intelligent transportation systems, and the Internet of Vehicles.
\end{IEEEbiography}

\begin{IEEEbiography}[{\includegraphics[width=1in,height=1.25in,clip,keepaspectratio]{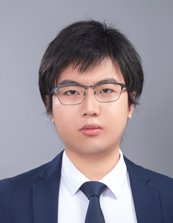}}]{Xincao Xu}
received the B.S. degree in network engineering from the North University of China, Taiyuan, China, in 2017. He is currently pursuing the Ph.D. degree in computer science at Chongqing University, Chongqing, China. His research interests include vehicular networks, edge computing, and deep reinforcement learning.
\end{IEEEbiography}

\begin{IEEEbiography}[{\includegraphics[width=1in,height=1.25in,clip,keepaspectratio]{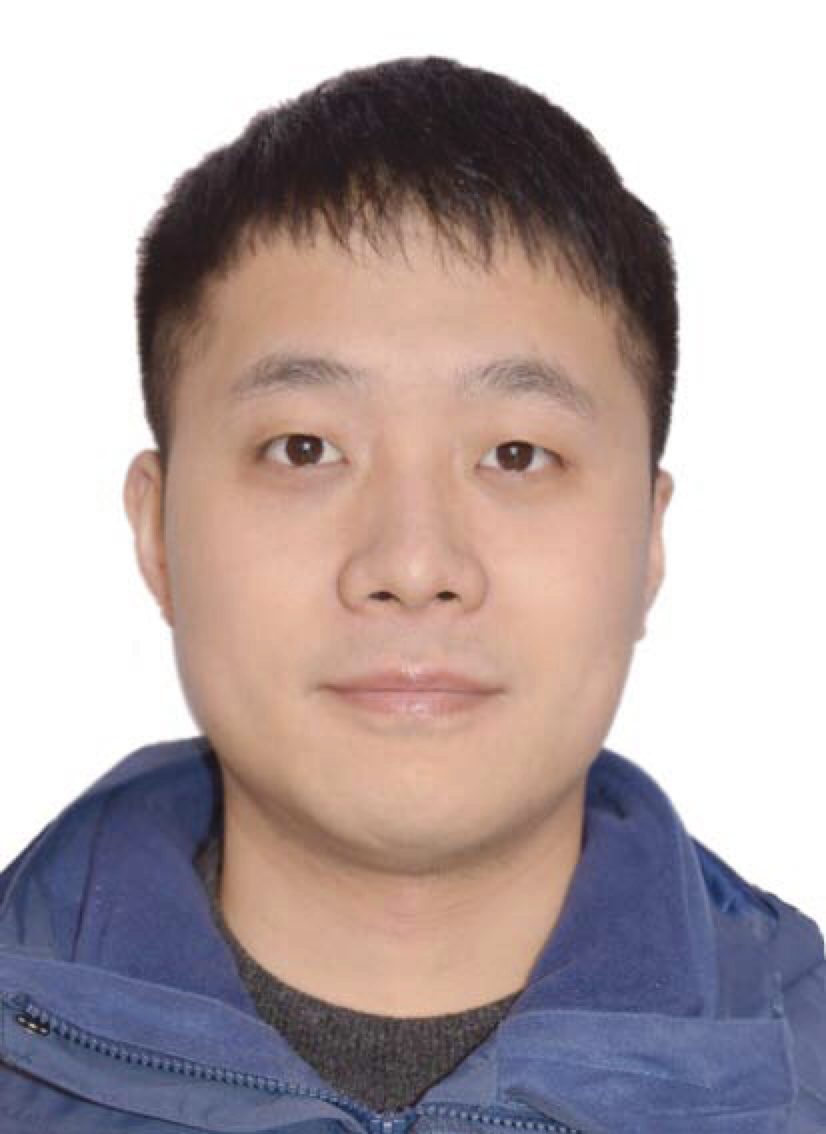}}]{Penglin Dai}
(Member, IEEE) received the B.S. degree in mathematics and applied mathematics and the Ph.D. degree in computer science from Chongqing University, Chongqing, China, in 2012 and 2017, respectively. He is currently an Associate Professor with the School of Computing and Artificial Intelligence, Southwest Jiaotong University, Chengdu, China. His research interests include intelligent transportation systems and vehicular cyber-physical systems.
\end{IEEEbiography}

\begin{IEEEbiography}[{\includegraphics[width=1in,height=1.25in,clip,keepaspectratio]{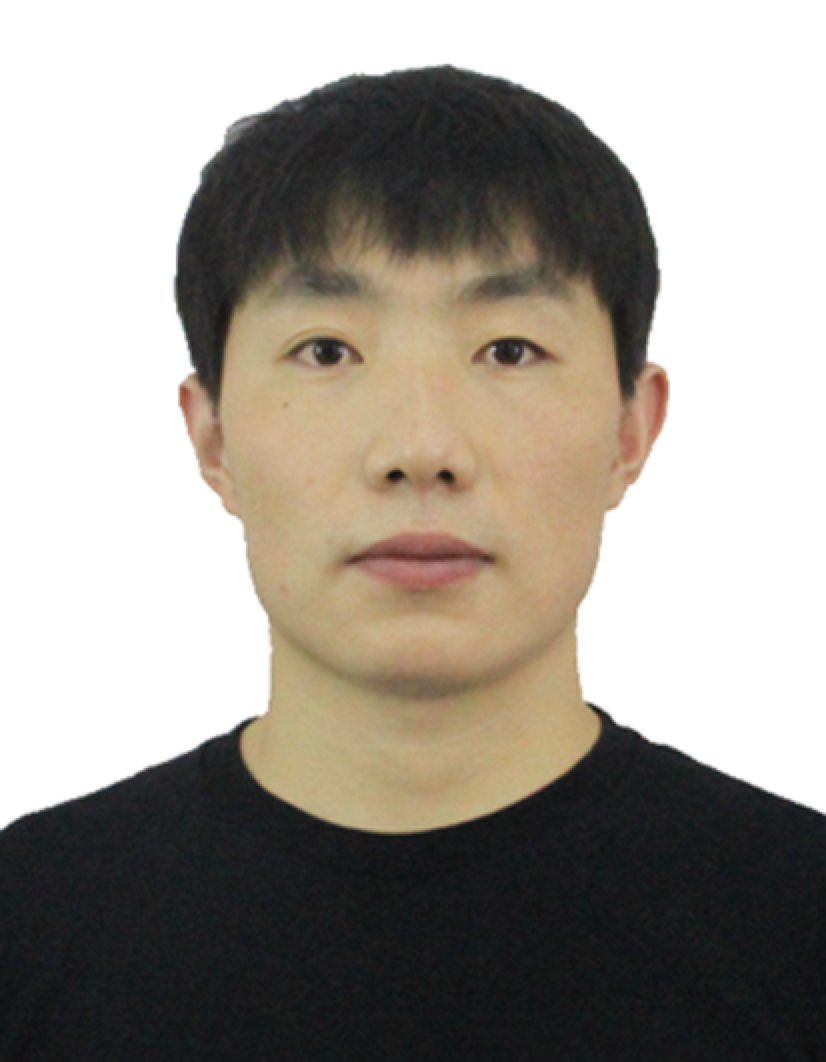}}]{Biwen Chen} received the Ph.D. degree from School of Computer, Wuhan University in 2020. He is currently an assistant professor of the School of Computer, Chongqing University. His main research interests include cryptography, information security and blockchain.
\end{IEEEbiography}

\vfill

\end{document}